\documentclass[12pt]{article}
\usepackage{color}
\usepackage{amsmath}
\usepackage{arxiv}
\usepackage{lyu}
\usepackage{xcolor}
\usepackage{pgfplots}
\usetikzlibrary{calc}
\pgfplotsset{compat=1.17}
\pgfplotsset{colormap={violet}{rgb255=(25,25,122) rgb255=(238,140,238) color=(white)}}
\usepackage[utf8]{inputenc} 
\usepackage[T1]{fontenc}    
\usepackage{hyperref}       
\usepackage{url}            
\usepackage{booktabs}       
\mathchardef\mhyphen="2D
\usepackage{amsfonts}       
\usepackage{nicefrac}       
\usepackage[bottom]{footmisc}

\usepackage{microtype}      
\usepackage{lipsum}
\usepackage{comment}
\usepackage{subcaption}
\usepackage{graphicx}
\usepackage{soul}           
\definecolor{myGreen}{rgb}{0.17254902, 0.62745098, 0.17254902}

\usepackage{algorithm}
\usepackage{algpseudocode}
\usepackage{appendix}
\usepackage{multirow}
\usepackage{amssymb}
\usepackage{amsthm}
\usepackage{amssymb}
\usepackage{mathabx}
\usepackage{tikz}
\usetikzlibrary{shapes,arrows}
\usetikzlibrary{positioning}
\usetikzlibrary{arrows,
    chains,
    decorations.markings,
    shadows, shapes.arrows,shapes, fit}
\newtheorem{theorem}{Theorem}[section]

\newtheorem{proposition}[theorem]{Proposition}

\newtheorem{definition}[theorem]{Definition}

\title{On the generalization ability of coarse-grained molecular dynamics models for non-equilibrium processes}
\renewcommand{\shorttitle}{}

\author{
 Liyao Lyu\\
 Department of Computational Mathematics, Science \& Engineering, Michigan State University, MI 48824, USA
\And
 Huan Lei \thanks{leihuan@msu.edu} \\ 
 Department of Computational Mathematics, Science \& Engineering, Michigan State University, MI 48824, USA\\
 Department of Statistics \& Probability, Michigan State University, MI 48824, USA}

\begin{document}

\maketitle

\keywords{} 
\begin{abstract}
One essential goal of constructing coarse-grained molecular dynamics (CGMD) models is to accurately predict non-equilibrium processes beyond the atomistic scale. While a CG model can be constructed by projecting the full dynamics onto a set of resolved variables, the dynamics of the CG variables can recover the full dynamics only
when the conditional distribution of the unresolved variables is close to the one associated with the particular projection operator. In particular, the model's applicability to various non-equilibrium processes is generally unwarranted due to the inconsistency in the conditional distribution. 
Here, we present a data-driven approach for constructing CGMD models that retain certain generalization ability for non-equilibrium processes. Unlike the conventional CG models based on pre-selected CG variables (e.g., the center of mass), the present CG model seeks a set of auxiliary CG variables based on the time-lagged independent component analysis to minimize the entropy contribution of the unresolved variables. This ensures the distribution of the unresolved variables under a broad range of non-equilibrium conditions approaches the one under equilibrium. Numerical results of a polymer melt system demonstrate the significance of this broadly-overlooked metric for the model's generalization ability, and the effectiveness of the present CG model for predicting the complex viscoelastic responses under various non-equilibrium flows. 
\end{abstract}

\section{Introduction}

Predictive modeling of the collective behaviors of multi-scale physical systems poses a persistent challenge for both fundamental science advancement and various applications \cite{More_Anderson_Science_1972}. While the canonical molecular dynamics enables us to faithfully account for the micro-scale interactions, numerical simulations often show limitations for systems without clear scale separation, where the computational cost could become prohibitive to reach the resolved scale of interest. This motivates the construction of various coarse-grained molecular dynamics (CGMD) models.  By picking a set of collective variables (CVs), the CG models seek the reduced dynamics with less degrees of freedom to achieve a broader range of spatio-temporal scales. The constructed CG models are generally governed by a conservative CG potential representing the free energy of the resolved CVs and a memory term (along with a coherent noise) representing the energy dissipation that arises from the coupling with unresolved variables. Extensive research efforts \cite{Tuckerman_adiabatic_JCP_2002,Eric_TAMD_CPL_2006,Izvekov_Voth_JPC_2005,noid_multiscale_2008,Rudd_APS_1998,Lyubartsev_PRE_1995,Shell_JCP_2008,Nielsen_JPCM_2004,Laio_Parrinello_PNAS_2002,Darve_Pohorille_JCP_2001,Soper_Chem_Phys_1996,Reith_JCP_2003,Nielsen_JPCM_2004,das2012multiscale}, including machine learning (ML) based approaches \cite{Behler_Parrinello_PRL_2007, Stecher_Thomas_JCTC_2014, John_ST_JPCB_2017, Lemke_Tobias_JCTC_2017, chmiela2017machine, Zhang_DeePCG_JCP_2018, Ge_Lei_JCP_2023, van2023hyperactive}, have been dedicated to developing the CG potential to preserve the marginal density distribution of the CG variables and hence the various static properties. To retain the CG dynamics, the memory term needs to be properly introduced. Several approaches have been developed based on the direct approximations \cite{Darve_PNAS_2009, Lei_Cas_2010, hijon2010mori, Yoshimoto_Kinefuchi_PRE_2013,hudson2020coarse} of the Mori-Zwanzig (MZ) projection formalism \cite{Mori1965, Zwanzig73} and data-driven parameterization \cite{lange2006collective, ceriotti2009langevin, coifman2008diffusion, crosskey2017atlas, baczewski2013numerical, MaLiLiu16, Lei_Li_PNAS_2016, russo2019deep, Jung_Hanke_JCTC_2017, Lee2019, Klippenstein_Vegt_JCP_2021, vroylandt2022likelihood, SheZ_JCP_2023, xie2023gle, Ge_Lei_GLE_PRL_2024} of empirical forms (e.g., dissipative particle dynamics \cite{Hoogerbrugge_SMH_1992, Espanol_SMO_1995}, the generalized Langevin dynamics \cite{Zwanzigbook}). A recent work \cite{lyu2023construction} proposed a symmetry-preserving representation of the memory term that accounts for both non-Markovianity and the many-body nature among the CG variables, which proves to be crucial for diffusion and transport processes on a resolved scale.

Ideally, by accurately constructing the conservative free energy and the memory term, the CG models will enable us to quantify the propagation of the micro-scale interactions and therefore probe the collective behaviors across multiple scales. However, the validity and generalization ability for practical applications remains under-explored. In particular, for extensive MD systems (i.e., the number of molecules proportional to the system size), the CG variables are often chosen \emph{a priori} such as the centers of mass (COMs) of individual molecules; the CG model is generally constructed such that certain dynamic properties (e.g, the mean square displacement, velocity correlation functions) under equilibrium can be properly reproduced. On the other hand, the model's applicability for non-equilibrium processes remains questionable. For instance, if we coarse grain a polymer melt system using the COMs of individual molecules, we should not expect the CG model can capture the visco-elastic responses arising from the molecule deformation under an external flow field.  From a model reduction perspective, this limitation arises from the choice of the CG projection operator, which is generally defined with respect to the (marginal) equilibrium density of the full MD system. Accordingly, reduced dynamics can recover the full MD prediction only when the underlying distribution is close to equilibrium. However, this caveat seems to be broadly overlooked in existing CGMD models that use equilibrium dynamic properties as the metric, and therefore poses fundamental challenges for predicting the non-equilibrium processes in real applications.

In principle, we may directly construct the reduced model for a non-equilibrium process by defining a projection operator with respect to a specific probability measure. However, the reduced dynamics typically relies on a time-dependent projection operator (e.g., see Ref. \cite{kawasaki1973theory,willis1974time,baxevani2023bottom}), which, as a result, generates a non-stationary memory term that may not be transferred to a different scenario. In this work, we present a new approach for constructing reliable CGMD models applicable to non-equilibrium processes. Rather than using pre-selected CG variables such as the COMs, we seek a set of auxiliary CG variables as the generalized coordinates of each molecule for the optimal representation of both the intra$\mhyphen$ and inter$\mhyphen$molecular interactions. The key observation is that by systematically introducing these auxiliary CG variables, the conditional distribution of unresolved variables under various non-equilibrium conditions approaches that under equilibrium conditions. This warrants the generalization ability and enables us to transfer the empirical approximation of the non-equilibrium processes into the construction of the Zwanzig's projection dynamics with respect to a probability density distribution with augmented conditional variables. To construct the reduced dynamics, we note that both the free energy and the memory term exhibit the many-body nature. We generalize the symmetry-preserving neural network representation of the state-dependent memory developed in our previous work \cite{lyu2023construction} to model both the intra$\mhyphen$ and inter$\mhyphen$energy dissipation. For each CG variable, we further introduce a number of non-Markovian features, allowing a coherent white noise term to be naturally imposed that satisfies the second fluctuation-dissipation theorem and preserves the invariant distribution. We demonstrate the proposed method by constructing the CGMD models of a polymer melt system. Numerical results show that the CG model based on the pre-selected COMs can recover the dynamic properties near equilibrium but is insufficient to predict the reduced dynamics under external flow conditions. Conversely, the present model guarantees that the unresolved orthogonal dynamics remains near equilibrium, ensuring the applicability of the projection formalism, and therefore, accurately predict non-equilibrium processes under various flow field conditions.

\section{Methods}
\label{sec:method}
\subsection{Preliminaries}
\label{sec:preliminary}
Let us consider a full MD system consisting of $M$ molecules with a total number of $N$ atoms. For simplicity, we assume each molecule consists of $N_m$ atoms, and the mass is set to be unit. The full model is governed by 
\begin{equation}
 \dot{\mb z} = \mb S\nabla H(\mb z) \quad \mb z(0) = \mb z_0, 
 \label{eq:full_model}
\end{equation}
where $H(\mb z)$ is the Hamiltonian, $\bz = \left[\bq^{1}, \bq^{2}, \cdots, \bq^{M}, \bp^{1},  \bp^{2}, \cdots, \bp^{M}\right]$ is the phase space vector and $\mb S$ is the symplectic matrix. Specifically, $\bq^I = \left[\bq^{I}_{1},\cdots,\bq^{I}_{N_m}], \bp^I = [\bp^{I}_{1},\cdots,\bp^{I}_{N_m}\right]\in \mathbb{R}^{N_m\times 3 }$ represents the position and momentum vectors of the $I\mhyphen$th molecule for $I = 1, 2, \cdots M$. In this work, the superscript indices in capital letter $I$ and $J$ ranging from $1$ to $M$ label the molecule index, while the subscript indices $n$ and $l$ ranging from $1$ to $N_m$ label the individual particles within the molecule. To construct the reduced model, we define the CG variables of each molecule $\mathbf Z^I = \left[\mathbf Q^I,\mathbf P^I\right]$ via a map $\phi:\mathbb{R}^{N_m\times 6 } \to \mathbb{R}^{m\times6}$, i.e., 
\begin{equation}
    \mathbf Q^I =  \phi^Q(\mathbf q^I) \quad \mathbf P^I = \phi^P(\mathbf q^I, \mathbf p^I),
\end{equation}
where $\mathbf Q^I \in \mathbb{R}^{m\times 3}$ and $\mathbf P^I \in \mathbb{R}^{m\times 3}$ represent the generalized coordinates and momenta of the $I\mhyphen$th molecule, respectively. Let $\mb Q = \left[\mb Q_1, \mb Q_2, \cdots, \mb Q_{mM}\right]$ and $\mb P = \left[\mb P_1, \mb P_2, \cdots, \mb P_{mM}\right]$ denote the CG variables of the full system, where $\mathbf Q^I = \left[\mb Q_{(I-1)m + 1}, \cdots, \mb Q_{Im}\right]$ with $\mathbf Q^I_j = \mb Q_\mu$ for $\mu={(I-1)m +j}$, and similar for $\mb P^I$. In this work, the Greek letters $\alpha,\beta,\mu,\nu$ ranging from $1,\cdots,mM$ label the global indices of CG coordinate. 

In particular, one natural choice is to define $\mb Q^I\in \mathbb{R}^{3}$ as the COM  and $\mb P^I\in \mathbb{R}^{3}$ the total momentum of the $I\mhyphen$th molecule (e.g., see Refs. \cite{Lei_Cas_2010, hijon2010mori, Yoshimoto_Kinefuchi_PRE_2013}), which essentially eliminates the intra-molecular DOFs. 
The dynamic evolution is governed by $\dot{\mb Z} = \mathcal{L} \mb Z$, where the Louville operator $\mathcal{L}\phi(\mb z) := -((\nabla H({\mb z}_0))^T \mb S\nabla {\mb z}_0)\phi(\mb z)$ depends on the full phase space vector $\mb z$. To derive the reduced dynamics in terms of $\mb Z(t)$, we define the Zwanzig's projection operator as a condition expectation with respect to $\mb Z(0) = \bm Z$, i.e., 
\begin{equation}
\mathcal P_{\bm Z} f(\mb z):= \mathbb{E}[f(\mb z) \vert \phi(\mb z) = \bm Z ] =  \int \delta(\mb\phi(\mathbf{z}) - \bm Z) \rho_{0}(\mathbf{z}) f(\mathbf{z}) \mathrm{d} \mathbf{z} / \int \delta(\mb\phi(\mathbf{z}) - \mb Z) \rho_{0}(\mathbf{z}) \mathrm{d} \mathbf{z},
\label{eq:projection_operator}
\end{equation}
where $\rho_0(\mb z) \propto \exp\left[-\beta H(\mb z)\right]$ represents the equilibrium Boltzmann distribution. Accordingly, we may project the evolution dynamics $\dot{\mb Z} = \mathcal{L} \mb Z$ on the sub-space of the CG variables. Using the Duhamel-Dyson formula, the reduced dynamics takes the form
\begin{equation}
\dot{\mb Z}(t) = {\rm e}^{\mathcal{L}t}\mathcal{P}_{\bm Z}\mathcal{L}\mb Z(0) + \int_0^t \diff s {\rm e}^{\mathcal{L}(t-s)}\mathcal{P}_{\bm Z} \mathcal{L} {\rm e}^{\mathcal{Q}_{\bm Z}\mathcal{L}s} \mathcal{Q}_{\bm Z}\mathcal{L}\mb Z(0)   + {\rm e}^{\mathcal{Q}_{\bm Z}\mathcal{L}t} \mathcal{Q}_{\bm Z}\mathcal{L}\mb Z(0), 
\label{eq:Z_MZ_full}
\end{equation}
where $\mathcal{Q}_{\bm Z} = \mb I - \mathcal{P}_{\bm Z}$. With some further assumptions, the reduced dynamics can be simplified into the following integro-differential equation 
\begin{equation}
\begin{split}
\dot{\mb Q} &= \mb M^{-1} \mb P \\
\dot{\mb P} &= -\nabla U(\mb Q) + \int_0^t \mb K(\mb Q(s), t-s)  \dot{\mb Q}(s) \diff s + \mb R(t), 
\end{split}
\label{eq:MZ_full}
\end{equation} 
where $\mb M$ is the mass matrix of the CG variables, $U(\mb Q)$ is the conservative free energy, and $\mb K(\mb Q, t)$ is the memory term assumed to be independent of $\mb P$ and $\mb R(t)$ is the fluctuation force. 

In principle, with proper construction of the individual modeling terms, Eq. \eqref{eq:MZ_full} provides an accurate CGMD model of the full dynamics \eqref{eq:full_model}. However, its applicability for practical applications, especially the generalization ability for modeling non-equilibrium processes remains questionable. This caveat is rooted in the definition of the projection operator in Eq. \eqref{eq:projection_operator}, where the conditional expectation is defined with respect to the marginal density of the equilibrium distribution $\rho_0(\mb z)$.  Accordingly, the derived reduced dynamics \eqref{eq:MZ_full} assumes that the initial distribution of the full model satisfies $\rho(\mb z) \propto \delta(\phi(\mb z) - \bm Z) \rho_0(\mb z)$. Hence, Eq. \eqref{eq:MZ_full} is valid only when the underlying distribution of the unresolved variables is close to equilibrium. Unfortunately, this condition is generally not guaranteed when we model non-equilibrium processes (e.g.,  full model \eqref{eq:full_model} in the presence of an external field) using the CG variables like the COMs of individual molecules, which poses a fundamental challenge for the transferability of the reduced model \eqref{eq:Z_MZ_full} and \eqref{eq:MZ_full} in real applications.

\subsection{Construction of the auxiliary CG variables}
\label{sec:auxiliary_CG}
Following the above discussion, let $H_e(\mb z)$ and $\rho_e(\mb z)$ denote the Hamiltonian and an initial (e.g., the steady-state) distribution under an external field. Theoretically, we may define the Zwanzig's projection operator with respect to $\rho_e(\mb z)$ and derive the reduced dynamics under this external field. However, the constructed reduced dynamics could be valid only for this specific condition but lack the generalization ability for other external field conditions. 

To address this issue, we transfer the exhausting effort of constructing specific reduced models for individual non-equilibrium processes to pursuing the following question: how to ensure the conditional projection operator defined by Eq. \eqref{eq:projection_operator} remains valid for a range of external conditions? A key observation is that by introducing a set of auxiliary CG variables $\mathbf{Z}^I = [ \mathbf{Z}^I_{1}, \mathbf{Z}^I_{2}, \cdots, \mathbf{Z}^I_{m}]$ for the individual molecules $I=1, 2, \cdots, M$, the conditional probability density function (PDF) under external field approaches that under the equilibrium distribution, i.e., 
\begin{equation}
\frac{1}{\mathcal{Z}_e}\prod_{I=1}^M \prod_{j=1}^m \delta({\phi(\mb z^I)}_{j} - {\bm Z}^I_{j}) \rho_e(\mb z) \approx \frac{1}{\mathcal{Z}_0} \prod_{I=1}^M \prod_{j=1}^m \delta({\phi(\mb z^I)}_{j} - {\bm Z}^I_{j}) \rho_0(\mb z).
\label{eq:conditional_prob_metric}
\end{equation}
Intuitively, this approximation can be understood as follows: As we increase the number of resolved CG variables, the entropy contribution of the unresolved ones decreases. Accordingly, the free energy of different external conditions approaches the equilibrium case. This is somewhat similar to the work \cite{Lei_Wu_E_2020,yu2005micro} on modeling the constitutive closure of polymer solution, where a set of generalized conformation tensors are introduced to represent the subgrid polymer configurations under various flow conditions.

To construct the auxiliary CG variables, we seek a linear map $\phi(\cdot)$ in the form of a matrix $W \in \mathbb{R}^{N_m \times m}$, i.e., 
\begin{equation}
    \phi^Q(\mathbf q^I)_j =  \frac{\sum_{n=1}^{N_m} w_{nj} \mathbf{q}^{I}_n}{\sum_{n=1}^{N_m} w_{nj}},  \quad 
    \phi^P(\mathbf p^I)_j = \sum_{i=1}^{m}M_{ji}\frac{\sum_{n=1}^{N_m} w_{ni} \mathbf{p}^{I}_n}{\sum_{n=1}^{N_m} w_{ni}},
\label{eq:CG_linear}
\end{equation}
where the mass matrix is defined as $M_{ji} = \sum_{n=1}^{N_m} w_{nj}w_{ni} $, and the normalization ensures that the transitional and rotational symmetry of the CG variables, i.e., 
\begin{equation}
\begin{split}
\phi(\mathbf q^I + \mathbf c , \mathbf p^I ) &=  [\phi^Q(\mathbf q^I+ \mathbf c),\phi^P(\mathbf p^I)] = [\phi^Q(\mathbf q^I)+ \mathbf c,\phi^P(\mathbf p^I)]  \\
\phi(\mathbf q^I \mathcal U, \mathbf p^I  \mathcal{U} ) &=  [\phi^Q(\mathbf q^I\mathcal{U}),\phi^P(\mathbf p^I \mathcal{U})] = [\phi^Q(\mathbf q^I)\mathcal{U},\phi^P(\mathbf p^I)\mathcal{U}],   \\
\nonumber
\end{split}
\end{equation}
for any $\mb c\in \mathbb{R}^3$ and $\mathcal{U} \in {\rm SO}(3)$. As shown below, Eq. \eqref{eq:CG_linear} can be loosely viewed as seeking the principal normal modes of the full dynamics. 
We note that it is possible to construct $\phi(\cdot)$ as a non-linear encoder to seek optimal CG representations of the inter$\mhyphen$ and intra$\mhyphen$ molecular interactions. In this work, we focus on developing a general CG framework of introducing auxiliary CG variables to enhance the generalization ability and examining Eq. \eqref{eq:conditional_prob_metric} as a valid metric for modeling non-equilibrium processes. We restrict $\phi(\cdot)$ to be linear; the model reduction in terms of nonlinear CG variables will be investigated in future work. 

Furthermore, we impose the following two constraints to the weight matrix $\mb W$, i.e., 
\begin{equation}
0 < w_{ni} < 1, \quad  \sum_{i=1}^m w_{ni} = 1 \quad \text{for all } n = 1,\cdots, N_m.
\label{eq:W_constraint}
\end{equation}
We emphasize that these constraints do not impose further restrictions on constructing the CG variables. Specifically, for any mapping defined by a matrix $\mb W_1 \in \mathbb{R}^{N_m \times (m-1)}$, there exists an equivalent mapping defined by a matrix $\mb W_2 \in \mathbb{R}^{N_m \times m}$ which satisfies the above constraints \eqref{eq:W_constraint}. We refer to the Appendix for the detailed discussion. 
Furthermore, we note that the second constraint ensures that the COM of the full molecule is consistent with that of the CG variables, i.e.,  $\sum_{j=1}^m \left(\sum_{n=1}^{N_m} w_{nj} \right) \mb Q^{I}_j  =  \sum_{j=1}^m \sum_{n=1}^{N_m} w_{nj} \mathbf{q}^{I}_n = \sum_{n=1}^{N_m} \mathbf{q}^{I}_n$, which enables us to establish a fair comparison of the different CG models on the same metric in Sec. \ref{sec:numerical_results}. Moreover, Newton's third law can be naturally imposed in Eq. \eqref{eq:thirdlaw}.

To learn CG mapping $\phi(\cdot)$, we aim to find a weight matrix $\mathbf W$ such that condition \eqref{eq:conditional_prob_metric} holds, which, unfortunately, cannot be easily checked \emph{a priori}. Alternatively, we propose to learn the CG variables with the longest relaxation of the time correlation function. Specifically, we collect training samples under an external field and seek $\mb W$ by solving the optimization problem
\begin{equation}
    \max_{\mb W} \frac{\sum_{\tau}\sum_{t=0}^{t_c} \sum_{I,j}^{M,m} \mathbf V^I_j(\tau) \mathbf V^I_j(\tau+t)^T}{\sum_{\tau}\sum_{I,j}^{M,m}  \mathbf V^I_j(\tau) \mathbf V^I_j(\tau)^T}
    \label{eq:CVloss}
\end{equation}
under constraint \eqref{eq:W_constraint}, 
where $\mathbf V^I_j = \frac{\sum_{n=1}^{N_m} w_{nj} \mathbf{p}^{I}_n}{\sum_{n=1}^{N_m} w_{nj}} $, $t_c$ is a hyperparameter representing the cut-off time when the correlation function decays to $0$ and $\tau$ is randomly selected from a long MD trajectory under the steady state. Intuitively, Eq. \eqref{eq:CVloss} yields the eigenmodes of the largest (negative) eigenvalue of the operator $\mathcal{L}$ under the linear approximation \cite{crommelin2011diffusion}. We note that this is somewhat similar to the time-lagged independent component analysis (TICA) \cite{perez2013identification,molgedey1994separation,schwantes2013improvements} and dynamic mode decomposition (DMD) \cite{schmid2010dynamic,chen2012variants,kutz2016dynamic} to identify
the dominant dynamics from the simulation data. On the other hand, the framework for enhancing the generalization ability of the CGMD model for non-equilibrium processes has not been pursued.

\begin{figure}
    \centering
    \includegraphics[width=1\textwidth]{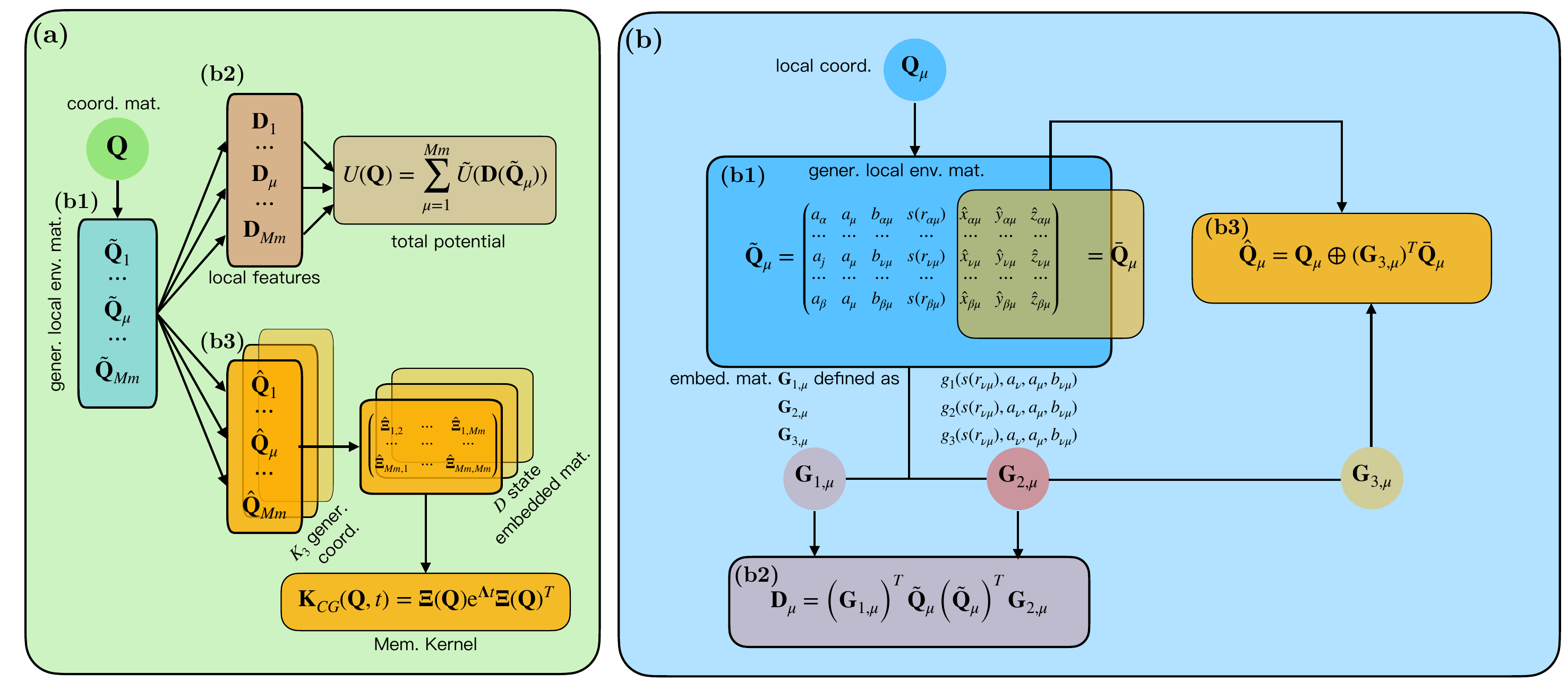}
    \caption{Diagram illustrating the neural network architecture for the CG conservative force and the memory kernel. (a) The construction of  CG potential function $U$ and memory kernel $\mathbf K$. Initially, $\mathbf Q$ is converted into a local environment matrix $\{\tilde {\mathbf Q}_\mu \}_{\mu=1}^{Mm}$. 
     Sub-networks, illustrated in (b), map $\tilde{ \mathbf Q}_\mu$ to a local feature $\mathbf D_\mu$ and generalized coordinate $\hat{\mb Q}_\mu$. Finally, the totel potential is constructed  by Eq. \eqref{eq:total-potential}, i.e.  $ U=  \sum_\mu^{Mm} \tilde{U}( D_\mu)$. The total memory kernel $\mathbf K$ is constructed with the state-dependent component of the memory kernel derived from $ \hat{\mathbf Q}_\mu$ using Eq. \eqref{eq:Xi_markovian} and the time-dependent component $\bm \Lambda$.  (b) The sub-networks map $\tilde{ \mathbf Q}_\mu$ to a local feature $D_\mu$ and generalized coordinate $ \hat{\mathbf Q}_\mu$.
     The neighbour of the $\mu\mhyphen$th CG coordinate is denoted by $\mathcal{N}_\mu= \{\alpha,\cdots,\nu,\cdots,\beta\}$. (b1) The $k\mhyphen$th row of generalized local environment matrix embeds the relative information between $\mu\mhyphen$th coordinate and its $k\mhyphen$th neighbor (labeled as the $\nu\mhyphen$th coordinate), including type embedding $a_\nu$, $a_\mu, b_{\nu\mu}$, and distance information. (b2) The $K_1\times K_2$ symmetry perserving feature  $\mathbf D_\mu$ is constructed by $(\mathbf G_{1,\mu})^T$, $\hat{\mathbf Q}_\mu$, $\hat{\mathbf Q}_\mu^ T$ and $\mathbf G_{2,\mu}$. (b3) The generalized coordinate is constructed from the local environment embedding matrix $\mathbf G_{3,\mu}$ and relative position $\bar {\mb Q}_\mu$, which is the last three columns of $\tilde{\mb Q}_\mu$.}
    \label{fig:mean_force}
\end{figure}

\subsection{Construction of the symmetry-preserving free energy function}
\label{sec:free_energy}

With the CG mapping $\phi$, the construction of the CG model proceeds with learning the free energy $U(\mb Q)$ and the memory kernel $\mb K(\mb Q,t)$.  To retain the extensive structure, the free energy $U(\mb Q)$ is decomposed into the local potential of individual CG coordinates, i.e., 
\begin{equation}
\label{eq:total-potential}
 U(\mb Q) = \sum_{\mu=1}^{M m}  \tilde{U}(\mb D(\tilde{\mb Q}_\mu)),    
\end{equation}
where $\tilde{U}(\cdot)$ represents the local potential, $\tilde{\mb Q}_\mu$ represents the local environment determined by the CG coordinates $\mb Q_\mu$ relative coordinates $\mb Q_\nu$ in its neighbors $\{\nu\left| \nu \in \mathcal{N}_{\mu}\right.\}$  and $\mb D(\cdot)$ represents the encoder functions that maps $\tilde{\mb Q}_\mu$ into symmetry-preserving features.  $ \mathcal{N}_{\mu}$
denotes the CG coordinate indices $\nu$ in the neighbour of CG coordinate indices $\mu$, such that $r_{\mu\nu} < r_c$. We define $N_\mu$ as the cardinality of the set  $ \mathcal{N}_{\mu}$.

Fig. \ref{fig:mean_force} sketches the neural network representation of $U(\mb Q)$, where a structure similar to Ref. \cite{Zhang_BookChap_NIPS_2018_v31_p4436} is used to preserve the translational and rotational symmetry constraints. To impose the permutational symmetry, we note that $U(\mb Q)$ remains invariant with the index permutation among the individual molecule, i.e., 
\begin{equation}
    U(\mb Q^{\sigma(1)}, \mb Q^{\sigma(2)}, \cdots, \mb Q^{\sigma(M)}) = U(\mb Q^{1}, \mb Q^{2}, \cdots, \mb Q^{M}),
\end{equation}
where $\left\{\sigma(I)\right\}_{I=1}^M$ represents an index permutation among moleculars. However, $U(\mb Q)$ does not necessarily remain invariant under an index permutation among the generalized CG coordinates, i.e.,
\begin{equation}
\begin{split}
U(\mb Q^{1}_1, \mb Q^{1}_2, \cdots, \mb Q^{2}_1,\mb Q^{2}_2\cdots ) &\neq U(\mb Q^{1}_1, \mb Q^{2}_2, \cdots, \mb Q^{2}_1, \mb Q^{1}_2\cdots), \\
U(\mb Q^{1}_1, \mb Q^{1}_2, \cdots, \mb Q^{2}_1,\mb Q^{2}_2\cdots ) &\neq U(\mb Q^{1}_2, \mb Q^{1}_1, \cdots, \mb Q^{2}_1, \mb Q^{1}_2\cdots), 
\end{split}
\end{equation}
where $\mb Q^I_i = \mb Q_{(I-1)m + i}$. This complication can not be remedied by introducing the particle type information as done in Ref. \cite{Zhang_BookChap_NIPS_2018_v31_p4436}. This is because, for a specific CG coordinate, its contribution to $U(\mb Q)$ could be either inter$\mhyphen$ or intra$\mhyphen$molecular interaction associated with other CG coordinates. Therefore, we introduce a parameter $\mb B_\mu\in \mathbb{R}^{N_\mu \times 1}$ to represent the type of interaction between the $\mu\mhyphen$th coordinate and its neighboring coordinates. Specifically, the $k\mhyphen$th row is set to be 1 if the $\mu\mhyphen$th coordinate and its $k\mhyphen$th neighbor (labeled as the $\nu\mhyphen$th coordinate) are part of the same molecule $I$, which means that there exist $i, j \in \{1, \cdots, m\}$ such that $\mu = (I-1)m + i$ and $\nu = (I-1)m + j$. Otherwise, it is set to be $-1$. In addition, we also introduce $\mb A_\mu\in \mathbb{R}^{N_\mu \times 2}$ to represent the type of the $\mu\mhyphen$th CG coordinate and the type of the CG coordinate within the neighborhood. The $k\mhyphen$th row of $\mb A_\mu$ is defined as $(a_\mu, a_\nu)$, where $a_\mu, a_\nu$ are set to be $i,j\in\{1,\cdots,m\}$ respectively.

To construct the symmetry-preserving local features, we define the local environment of the $\mu\mhyphen$th coordinate by $\tilde{\mb Q}_\mu \in \mathbb{R}^{N_\mu \times 7}$. The $k\mhyphen$th row of $\tilde{\mb Q}_\mu$ is constructed by the relative position between the $\mu\mhyphen$th coordinate and the $k\mhyphen$th neighbor coordinate (labeled as the $\nu\mhyphen$th coordinate), i.e., $\tilde{\mb Q}_{\nu\mu}:=(a_\nu,a_\mu,b_{\nu\mu}, s(r_{\nu\mu}), \hat{x}_{\nu\mu}, \hat{y}_{\nu\mu}, \hat{z}_{\nu\mu})$, where $\mb r_{\nu\mu} = (x_{\nu\mu}, y_{\nu\mu}, z_{\nu\mu})$ denotes the relative position to its neighbor, $\hat{x}_{\nu\mu} = s(r_{\nu\mu})x_{\nu\mu}/r_{\nu\mu}$, and similar for $\hat{y}_{\nu\mu}$ and $ \hat{z}_{\nu\mu}$.  $s(r)$ is defined by
\[ s(r) = \begin{cases} 
\frac{r}{r_c^2} & \text{if } I = J, \\
\frac{1}{r} & \text{if } r < r_{cs} \text{ and } I \neq J ,\\
\frac{1}{2r} \left[ 1 + \cos\left (\frac{\pi (r - r_{cs})}{r_{c} - r_{cs}} \right) \right] & \text{if } r_{cs} \leq r < r_{c} \text{ and } I \neq J, \\
0 & {\rm else}\\ 
\end{cases} \]
where $I$ and $J$ represent the molecule indices of the $\mu\mhyphen$th and its neighbor coordinate. In particular, if the two coordinates belong to the same molecule, their intra-molecular interactions will always be accounted for. Otherwise, $s(r)$ is a smooth differentiable function that decays to 0 beyond $r_c$. $r_{cs}$ is a hyper-parameter and is set to be $r_c-1$ in this study. 

With the local environment $\tilde{\mb Q}_\mu$, we construct the local features  $\mb D \in \mathbb{R}^{K_1 \times K_2}$ by
\begin{equation}
\begin{split}
\mb D(\tilde{\mb Q}_\mu) 
&= \left(\mb G_{1,\mu}\right)^T {\tilde{\mb Q}}_\mu  \left({\tilde{\mb Q}}_\mu\right)^T \mb G_{2,\mu} \\ 
&:= \left(\sum_{\nu\in \mathcal{N}_{\mu}} \mb g_{1}(s(r_{\nu\mu}),a_\nu,a_\mu,b_{\nu\mu}) \tilde{\mb Q}_{\nu\mu} \right)\left(\sum_{\nu\in \mathcal{N}_{\mu}} \mb g_{2}(s(r_{\nu\mu}),a_\nu,a_\mu,b_{\nu\mu}) \tilde{\mb Q}_{\nu\mu}\right)^T,        
\end{split}
\end{equation}
where $\mb G_{1,\mu}\in \mathbb R ^{N_\mu \times K_1}, \mb G_{2,\mu}\in \mathbb R^{N_\mu \times K_2}$. The $k\mhyphen$th row of $\mb G_{1,\mu}$ is represented as a neural network $\mb g_{1}(s(r_{\nu\mu}),a_\nu,a_\mu,b_{\nu\mu})$ that embeds the relative information between the $\mu\mhyphen$th coordinate and its $k\mhyphen$th neighbor (labeled as the $\nu\mhyphen$th coordinate) in $\mathbb{R}^4$ to a feature space in $\mathbb{R}^{K_1}$. Similarly, $\mb G_{2,\mu}$ is represented by a neural network $\mb g_{2}(s(r_{\nu\mu}),a_\nu,a_\mu,b_{\nu\mu})$ mapping from $\mathbb{R}^4$ to $\mathbb{R}^{K_2}$.

To construct the free energy function $U(\mb Q)$, we conduct restrained MD simulations (see Sec. \ref{sec:memory} for details) and train the neural network representations $\tilde{U}(\cdot)$, $\mb g_1(\cdot)$ and $\mb g_2(\cdot)$ by minimizing the empirical loss between the MD and CG model, i.e., 
\begin{equation}
L_U = \sum_{s=1}^S  \left\Vert \nabla U(\mb Q^{(s)}) + \mathcal{F}(\mb Q^{(s)})\right\Vert^2,   
\end{equation}
where the superscript $s$ represents various configurations and the $\mathcal{F}$ represents the conservative force term sampled from the full MD simulation.

\subsection{Construction of the symmetry-preserving memory function}
\label{sec:memory}
Besides the free energy function $U(\mb Q)$, the reduced dynamics further depends on the memory function $\mb K(\mb Q, t)$. In particular, recent work \cite{lyu2023construction} shows that the memory function could exhibit a strong many-body nature. Empirical approximations such as the standard GLE with a homogeneous kernel and the dissipative particle dynamics with a pairwise decomposition generally show limitations in modeling the heterogeneous energy dissipation among the CG particles. To predict the collective dynamics, it is crucial to accurately model both the non-Markovian and many-body nature of the memory function. 

Following the neural network structure proposed in Ref. \cite{lyu2023construction}, we encode the many-body dissipative interactions among the CG particles by introducing coordinate features $\widehat{\mb Q}_\mu = [\widehat{\mb Q}_\mu^1, \widehat{\mb Q}_\mu^2, \cdots, \widehat{\mb Q}_\mu^{K_3}]\in \mathbb{R}^{K_3\times 3}$, where 
the feature $\widehat{\mb Q}_\mu$ is defined by
\begin{equation}
\begin{split}
\widehat{\mb Q}_\mu &= \mb Q_\mu \oplus  \left( \mb G_{3,\mu}\right)^T  \bar{\mb Q}_\mu \\
&:= \mb Q_\mu \oplus \sum_{\nu\in \mathcal{N}_{\mu} } \mb g_{3}(s(r_{\nu\mu}),a_\nu,a_\mu,b_{\nu\mu})  \bar{\mb Q}_{\nu\mu}, 
\end{split}
\end{equation}
where $\mb G_{3,\mu} \in \mathbb{R}^{N_\mu\times K_3}$, where $\mb g_3:\mathbb{R}^4\to \mathbb{R}^{K_3}$ is a neural network that encodes the dissipative interactions beyond the pairwise form. $\bar {\mb Q}_\mu \in \mathbb R^{N_\mu\times 3}$ denotes the
relative position between the $\mu\mhyphen$th coordinate and its local neighbor, whose $k\mhyphen$th row (corresponding to $\nu\mhyphen$th coordinate) is defined as $\bar {\mb Q}_{\nu\mu} := (\hat{x}_{\nu\mu}, \hat{y}_{\nu\mu}, \hat{z}_{\nu\mu})$ .
Accordingly, we can construct the state-dependent friction tensor in form $\bm\Gamma(\mb Q) =  \bm\Xi(\mb Q) \bm\Xi(\mb Q)^T$, where $\bm\Xi \in \mathbb{R}^{3mM \times 3mM}$. The entries $\bm\Xi_{\mu\nu}$ represents the friction between the $\mu\mhyphen$th and $\nu\mhyphen$th CG coordinates and takes the form
\begin{equation}
\begin{split}
\bm\Xi_{\mu\nu} = \sum_{k=1}^{K_3} h_k(\hat{\mb Q}_{\mu\nu} \hat{\mb Q}_{\mu\nu}^T  ) \hat{\mb Q}_{\mu\nu}^k \otimes \hat{\mb Q}_{\mu\nu}^k + h_0(\hat{\mb Q}_{\mu\nu} \hat{\mb Q}_{\mu\nu}^T ) \mb I, 
\end{split}
\label{eq:Xi_markovian}
\end{equation}
where $h: \mathbb{R}^{K_3\times K_3}\to \mathbb{R}^{K_3+1}$ are encoder functions to be learned. Furthermore, since the CG mapping defined by Eqs. \eqref{eq:CG_linear} and \eqref{eq:W_constraint} ensures that the COM of the CG coordinates of each molecule is the same as that of the full molecule, we can impose Newton's third law by choosing $\bm\Xi_{\mu\mu} = -\sum_{\nu\in \mathcal{N}_\mu} \bm \Xi_{\mu\nu}$. 

We can show that $\bm\Xi$ in the form of Eq. \eqref{eq:Xi_markovian} strictly preserve the translational, rotational, and permutational symmetry constraints, i.e., 
\begin{equation}
\begin{split}
\bm\Xi_{\mu\nu}(\mb Q_1 +\mb b, \cdots ,\mb Q_{mM} +\mb b) &= \bm\Xi_{\mu\nu}(\mb Q_1, \cdots,\mb Q_{mM}) \\ 
\bm\Xi_{\mu\nu}(\mb Q_1 \mathcal{U}, \cdots ,\mb Q_{mM} \mathcal{U}) &= \mathcal{U} \bm\Xi_{\mu\nu}(\mb Q_1, \cdots,\mb Q_{mM}) \mathcal{U}^T \\
\bm\Xi_{\mu\nu}(\mb Q^{\sigma(1)}, \mb Q^{\sigma(2)}, \cdots, \mb Q^{\sigma(M)}) &= \bm\Xi_{\sigma(\mu) \sigma(\nu)}(\mb Q^{1}, \mb Q^{2}, \cdots, \mb Q^{M})    
\end{split}
\label{eq:memory_symmetry_condition}
\end{equation}
for any $\mb c\in \mathbb{R}^3$, $\mathcal{U} \in {\rm SO}(3)$, and index permutation $\left\{\sigma(I)\right\}_{I=1}^M$, where $\mathbf Q^I = \left[\mb Q_{(I-1)m + 1}, \cdots, \mb Q_{Im}\right]$. 

To further model the non-Markovian effect, we seek a set of non-Markovian features $\bm \zeta := \left[\bm\zeta_{1}, \bm\zeta_{2}, \cdots, \bm\zeta_{D}\right]$ and learn the joint dynamics of $\left [\mb Q, \mb P, \bm \zeta\right]$ by modeling the friction tensor between $\mb P$ and  $\bm \zeta$ in form of Eq. \eqref{eq:Xi_markovian}, i.e.,
\begin{equation}
\begin{split}
 \dot{\mb Q} &= \mb M^{-1} \mb P \\ 
\dot{\mb P} &= -\nabla U(\mb Q) + \bm \Xi(\mb Q)\bm\zeta \\
\dot{\bm\zeta} &= - \bm \Xi(\mb Q)^T \mb V - \bm\Lambda \bm\zeta + \bm \xi(t),
\end{split}
\label{eq:CGMD}
\end{equation}
where $\bm\Xi = \left[\bm \Xi^1 \bm \Xi^2 \cdots \bm \Xi^D\right]$ and sub-matrix takes the form
\begin{equation}
\begin{split}
\bm\Xi^d_{\mu\nu} = \sum_{k=1}^{K_3} h^d_k(\hat{\mb Q}_{\mu\nu} \hat{\mb Q}_{\mu\nu}^T  ) \hat{\mb Q}_{\mu\nu}^k \otimes \hat{\mb Q}_{\mu\nu}^k + h^d_0(\hat{\mb Q}_{\mu\nu} \hat{\mb Q}_{\mu\nu}^T ) \mb I, 
\end{split}
\label{eq:Xi_nonmarkovian}
\end{equation}
where $h: \mathbb{R}^{K_3\times K_3}\to \mathbb{R}^{D(K_3+1)}$ and $d=1,\cdots, D$. $\bm\Lambda$ takes an extendable form $\bm\Lambda = \hat{\bm \Lambda}\otimes \mb I$, where $\mb I \in \mathbb{R}^{3M\times 3M}$ is the identity matrix and $\hat{\bm \Lambda} \in \mathbb{R}^{D\times D}$ specifies the coupling among the features. It takes the form $\hat{\bm \Lambda} = \hat{\mb L}\hat{\mb L}^T + \hat{\mb L}^a$, where $\hat{\mb L}$ is a lower triangular matrix and $\hat{\mb L}^a$ is an anti-symmetry matrix; it satisfies the Lyapunov stability condition $\hat{\bm \Lambda} + \hat{\bm \Lambda}^T \ge 0$.  Accordingly, we can introduce white noise term following $\left\langle  \bm \xi(t) \bm \xi(t') \right\rangle = \beta^{-1}(\bm\Lambda + \bm\Lambda^T) \delta(t-t')$. With this choice, we can show that Eq. \eqref{eq:CGMD} satisfies the second fluctuation-dissipation theorem.  We refer to Ref. \cite{lyu2023construction} for the detailed proof.  

To construct $\mb K(\mb Q, t)$, we learn the memory embedded in Eq. \eqref{eq:CGMD} in the form of $\bm\Xi(\mb Q(t)){\rm e}^{\bm\Lambda(t-s)}\bm\Xi(\mb Q(s))^T$ directly from the Zwanzig's formalism so that the many-body nature can be naturally inherited. Following Ref. \cite{hijon2010mori}, we use the restrained dynamics $\tilde{\mb z}(t)= {\rm e}^{\mathcal{R}t} \mb z(0)$ to approximate the orthogonal dynamics ${\rm e}^{\mathcal{Q}_{\bm Z}Lt}$ based on the observation $\mathcal{P}\mathcal{Q} = \mathcal{P}\mathcal{R}\equiv 0$. 
Accordingly, the memory term of Zwanzig's form (see Eq. \eqref{eq:Z_MZ_full}) can be approximated by  $\mb K_{MZ}(\mb Q, t) = \mathcal{P}_{\bm Z}\left[({\rm e}^{\mathcal{R} t}\mathcal{Q}_{\bm Z} \mathcal{L}\mb P) (\mathcal{Q}_{\bm Z} \mathcal{L}\mb P)^T\right]$ and the memory of the CG model reduces to $\mb K_{CG}(\mb Q, t) = \bm\Xi(\mb Q){\rm e}^{\bm\Lambda t}\bm\Xi(\mb Q)^T$.  To collect the training samples, we establish the restrained dynamics following 
\begin{equation}
\begin{split}
 \dot {\mathbf  q}^I &= \mathbf  p^I -  \mb W (\mb W^T \mb W) ^{-1} \mb W^T \mathbf p^I\\
 \dot {\mathbf  p}^I &= \mathbf  f^I -  \mb W (\mb W^T \mb W) ^{-1} \mb W^T \mathbf f^I\\
\end{split}
\label{eq:restraint}
\end{equation}
where $\mb f^I = \left[\mb f ^{I}_{1},\cdots,\mb f ^{I}_{N_m}\right]\in \mathbb{R}^{N_m\times 3 }$ is the force of the individual atoms belonging to the molecule $I$. The total force of each CG variable is defined as $\mathbf F^I_j  = \sum_{n=1}^{N_m} w_{nj} \mathbf f^I_n$, which is equivalently denoted by $\mb F_\mu$ with $\mu = (I-1)m+j$ and $\mu = 1,\cdots, mM$.

Under the restrained dynamics \eqref{eq:restraint}, the orthogonal fluctuation term corresponds to the random force of each CG variable, i.e.,  $\mathcal{Q}_{\bm Z} \mathcal{L}\mb P = \left[\delta \mb F_{1}, \delta \mb F_{2}, \cdots, \delta \mb F_{mM}\right]$, 
where 
$\delta \mb F_{\mu} = \mathbf F_{\mu}  - \left\langle \mathbf F_{\mu}\right\rangle $ 
represents the random force of the $\mu\mhyphen$th CG coordinate and $\left\langle \cdot \right\rangle$ represents the condition expectation whose generalization ability will be examined in Sec. \ref{sec:numerical_results}. Due to the constraint \eqref{eq:W_constraint}, the Newton's third law is naturally satisfied, i.e., 
\begin{equation}
    \begin{split}
        \sum_{\mu}^{mM}\mathbf F_\mu= \sum_{I,i}^{M,m} \sum_{n}^{N_m} w_{ni} \mathbf f^I_n  = \sum_{I,i}^{M,m} \sum_{n}^{N_m} w_{ni} \sum_{J,l}^{M,N_m}\mathbf f_{nl}^{IJ} 
         = \sum_{I,n}^{M,N_m}\sum_{J,l}^{M,N_m}\mathbf f_{nl}^{IJ} \equiv 0,
    \end{split}
    \label{eq:thirdlaw}
\end{equation}
where $\mathbf f^{IJ}_{nl}$ represents the force 
between the $n\mhyphen$th atom of the $I\mhyphen$th molecule and the $l\mhyphen$th atom of the $J\mhyphen$th molecule. Therefore, $\delta \mathbf F_\mu = -\sum_{\nu\neq \mu}\delta \mathbf F_\nu$ and $\bm\Xi_{\mu\mu} = -\sum_{\nu\neq \mu} \bm\Xi_{\mu\nu}$.

We estimate the memory term $\mathbf K_{MZ}(\mb Q, t)$ for each CG configuration. This enables us to train the memory function in terms of the encoders  $\left\{\mb g_3(\cdot), h\right\}$ and matrices $\{\hat{\mb L}, \hat{\mb L}^a\}$ by minimizing the empirical loss 
\begin{equation}
L_{K} = \sum_{s=1}^{N_s} \sum_{j=1}^{N_t} \left\Vert  \mb K_{CG}(\mb Q^{(s)}, t_j) -  \mb K_{MZ}(\mb Q^{(s)}, t_j)\right\Vert^2,  
\end{equation}
where $s$ represents the different CG configurations.

\section{Numerical Results}
\label{sec:numerical_results}

\subsection{The full atomistic model}
\label{sec:full_model}
To validate our method, we consider a full micro-scale model of a polymer melt system. Each polymer molecule consists of $N_m = 106$ atoms. The central atom is connected to 3 arms and each arm consists of $5$ atoms. At the end of each arm, there are $6$ additional arms connected to the end atom and each additional arm consists of $5$ atoms. The pairwise atomistic interaction is chosen to be the Weeks-Chandler-Andersen potential 
\begin{equation}
V_p(r) = \begin{cases} V_{\rm LJ}(r) - V_{\rm LJ}(r_c), ~r < R_c  \\
0, ~r \ge R_c 
\end{cases} \quad
\quad V_{\rm LJ}(r) = 4\epsilon \left[\left(\frac{\sigma}{r}\right)^{12} -  \left(\frac{\sigma}{r}\right)^6\right],
\label{eq:LJpotential}
\end{equation}
where $\epsilon = 1.0$ is the dispersion energy, $r_c = 2^{1/6}\sigma$ 
and $\sigma = 2.415$ is the hardcore distance.
The bond interaction is chosen to be the FENE potential
\begin{equation}
    V_b(r) =  -0.5KR_0^2 \ln\left[1-\left(\frac{r}{R_0}\right)^2\right],
\end{equation}
where stiffness is set to be  $K= 1$ and the distance is to be $R_0=5$.
The complete system consists of $M = 300$ molecules in a $90\times 90\times 90$ domain in the reduced unit. The periodic boundary condition is imposed along each direction. The system is equilibrated under the Nos\'e-Hoover thermostat with temperature $k_BT = 4.0$. In this work, the external force field is imposed on the $x\mhyphen y$ plane and the temperature is defined by the velocity in the $z$ direction. To collect the training samples, we impose an external force field $f_x$ along the $x\mhyphen$direction to generate the reverse Poiseuille flow, i.e., 
\begin{equation}
f_{x}(y) = \begin{cases}
f_0 & \text{if } 0<y<45,\\
-f_0 & \text{else,}
\end{cases}
\label{eq:external_force}
\end{equation}
where $f_0 = 0.01$ is the force magnitude. The full model is simulated for $1\times 10^6$ steps with a time step $\intd t=1\times 10^{-3}$  to achieve the steady state, followed by $5\times 10^5$ steps to collect the statistical samples in Eq \eqref{eq:CVloss}, with $t_c$ set to be $5$. To construct the CG models, we learn $m=3$ and $m=4$ CVs from the training set.  
Throughout the rest of this paper, we will denote the reduced model that only uses the standard COMs by CGCOM, and the reduced models with $3$ and $4$ auxiliary CVs (per each molecule) by CG3 and CG4, respectively. In particular, for the CG3 model, the obtained $3$ CG variables are equivalent which uniformly divides the individual molecule into three parts. In contrast, the obtained CG variables are inequivalent for the CG4 model. 

\subsection{Generalization ability of the CG models}
\label{sec:generalization_CG}

As discussed in Sec. \ref{sec:preliminary} and Sec. \ref{sec:auxiliary_CG}, the generalization ability of the CG models for the non-equilibrium processes relies on the assumption that the conditional PDF (i.e., $\propto \delta(\phi^Q(\mb q) - \mb Q) \rho_e(\mb q)$) used for defining the CG projection operator \eqref{eq:projection_operator} remains nearly the same, i.e., Eq. \eqref{eq:conditional_prob_metric} holds for various external flow conditions. 
In practice, the direct evaluation of this high-dimensional PDF becomes computationally intractable. 
Instead, we relax this metric by examining the second moments of the atomistic particle distribution conditional to fixed CG variables under various non-equilibrium conditions. Specifically, we examine the variation of the particle position in the orthogonal direction defined by $\mb q^I_{\perp, \mb W}= \mathbf q^I -  \mb W ( \mb W^T  \mb W)^{-1}  \mb W^T\mathbf q^I $, where $\mb q^I \in \mathbb{R}^{N_m \times 3}$ represents the atomistic positions of the $I\mhyphen$th molecule and $\mb W$ is the weight matrix constructed by Eq. \eqref{eq:CVloss}. 
The second-moment covariance matrix is defined by 
$$
\mathbf C(y) = \mathbb E^\text{neq}\left[\mb q^{I}_{\perp, \mb W} {\mb q^{I}_{\perp,  \mb W}}^T \mid y_c^{I}=y\right],
$$
where $\mathbf C(y) \in \mathbb{R}^{N_m\times N_m}$ and the conditional expectation $\mathbb E^\text{neq}$ is evaluated over the full molecules $I = 1, 2, \cdots, M$ under the stationary distribution generated by the external force field \eqref{eq:external_force}. The condition $y_c^I = y$ restricts the COMs of the sampling molecule along the $y\mhyphen$direction.

In particular, the external force generates the reverse Poiseuille flow along the $x\mhyphen$direction, where the flow shear rate $\frac{{\partial } u_x}{{\partial} y}$ and the external stress vary along the $y\mhyphen$direction. Specifically, the shear rate is close to $0$ at $y = L/4$ and $y = 3L/4$, and achieves the largest value at $y = 0$, $y=L/2$, and $y=L$. Therefore, the variation of $\mb C(y)$ with respect to $y$ provides a \emph{computationally accessible metric} for Eq. \eqref{eq:conditional_prob_metric}. If the conditional PDF is close to the equilibrium distribution, $\mb C(y)$ should be homogeneous. Conversely, the large variation implies the deviation from the equilibrium distribution.

\begin{figure}
    \centering
    \includegraphics[width=0.3\textwidth]{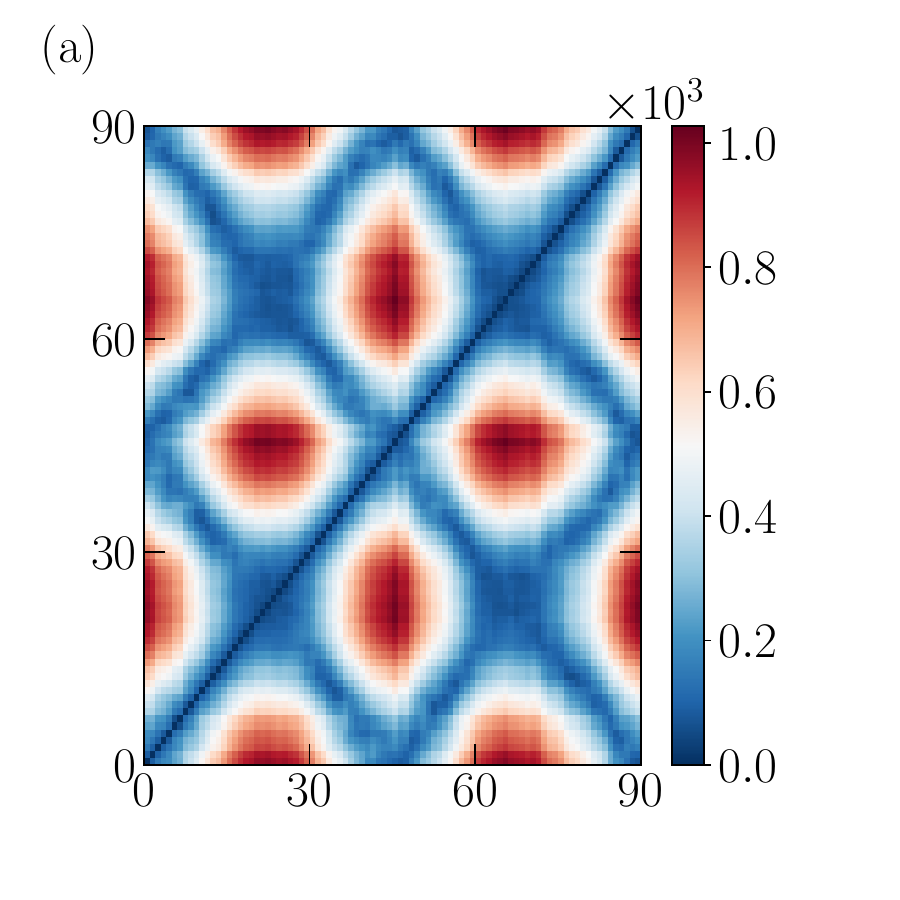}
    \includegraphics[width=0.3\textwidth]{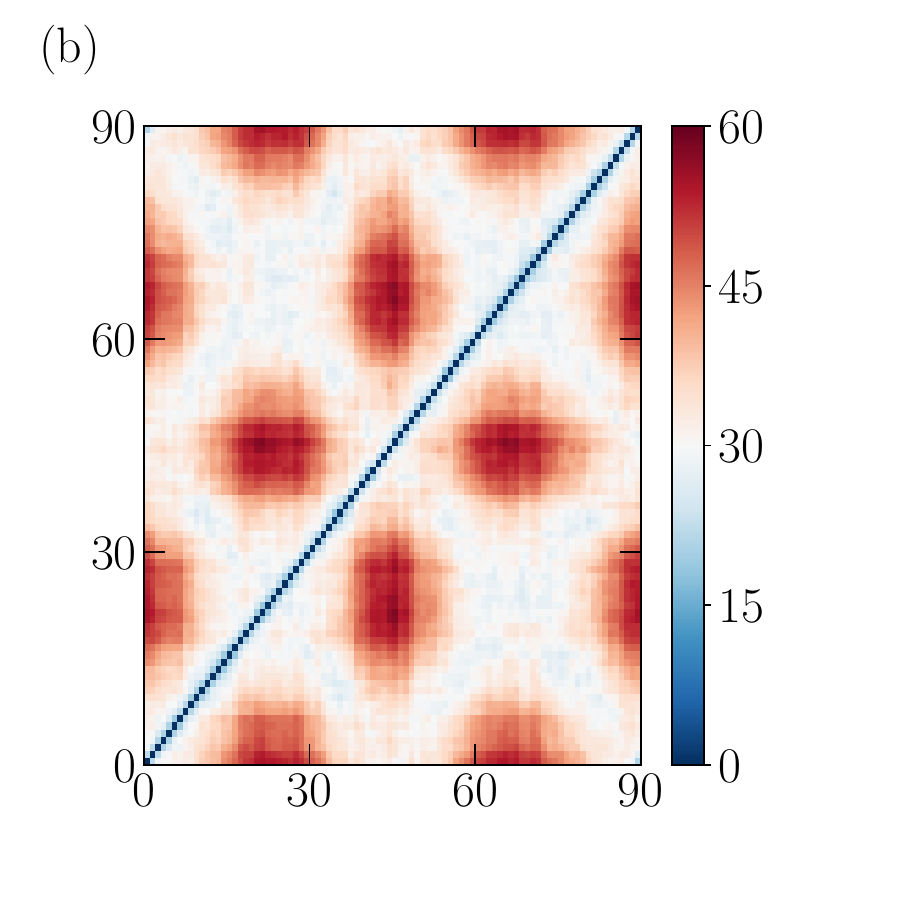}
    \includegraphics[width=0.3\textwidth]{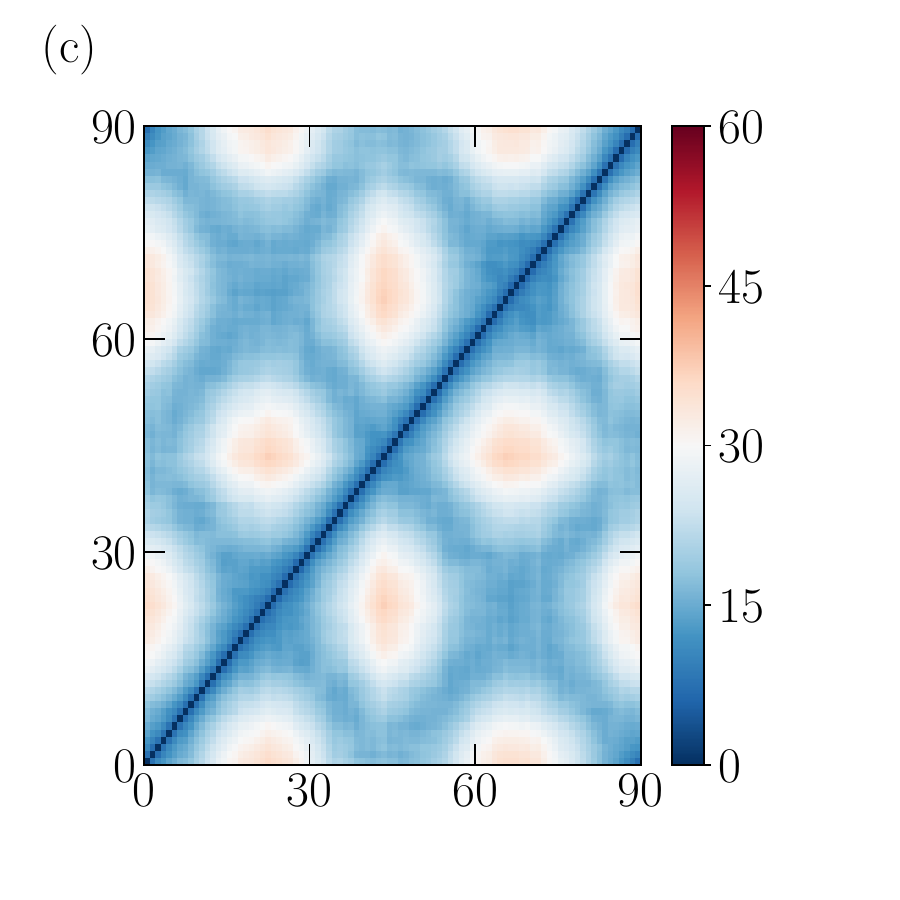}
    \includegraphics[width=0.3\textwidth]{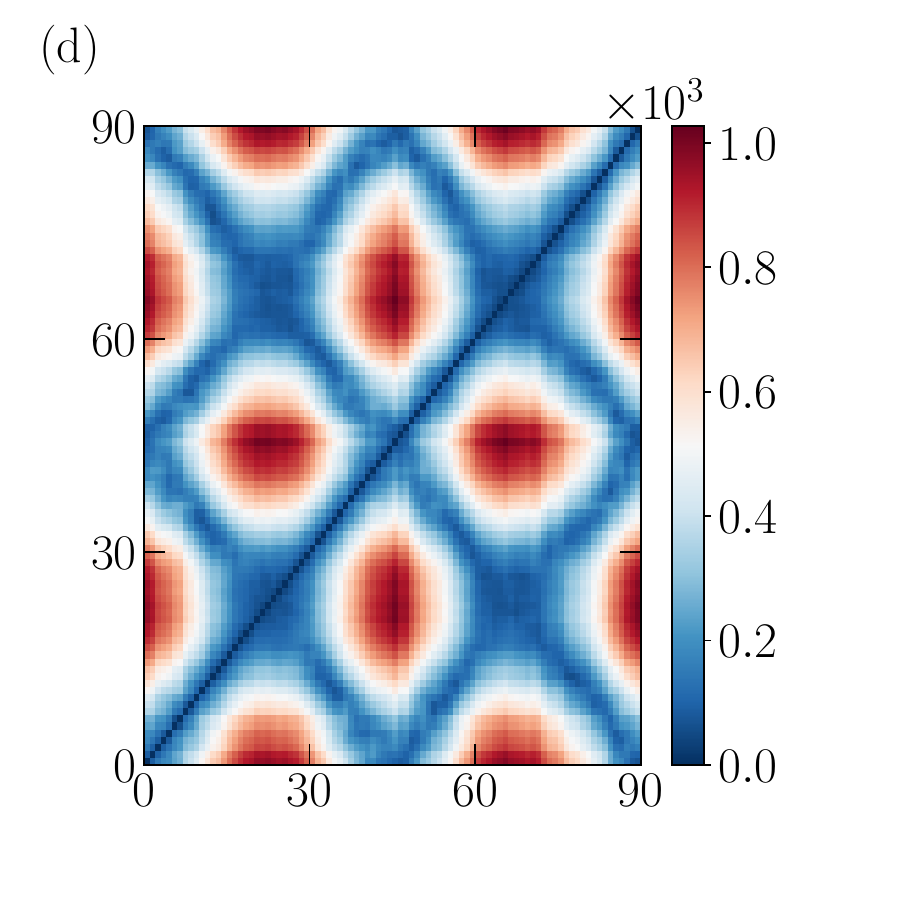}
    \includegraphics[width=0.3\textwidth]{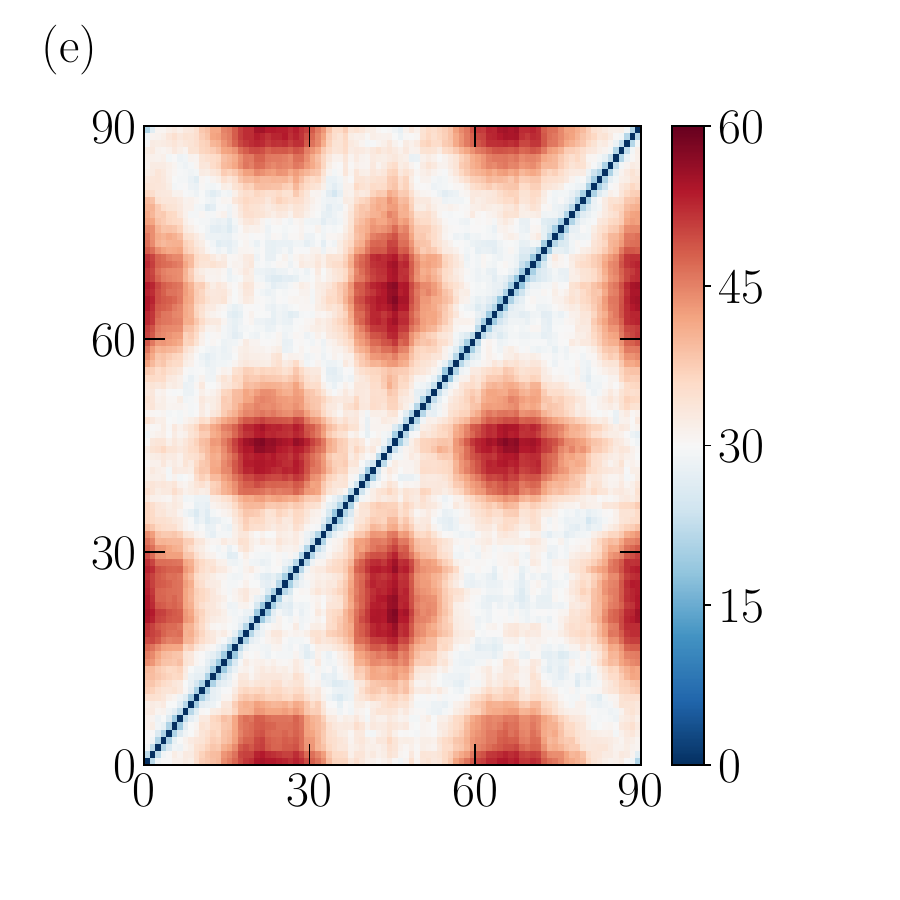}
    \includegraphics[width=0.3\textwidth]{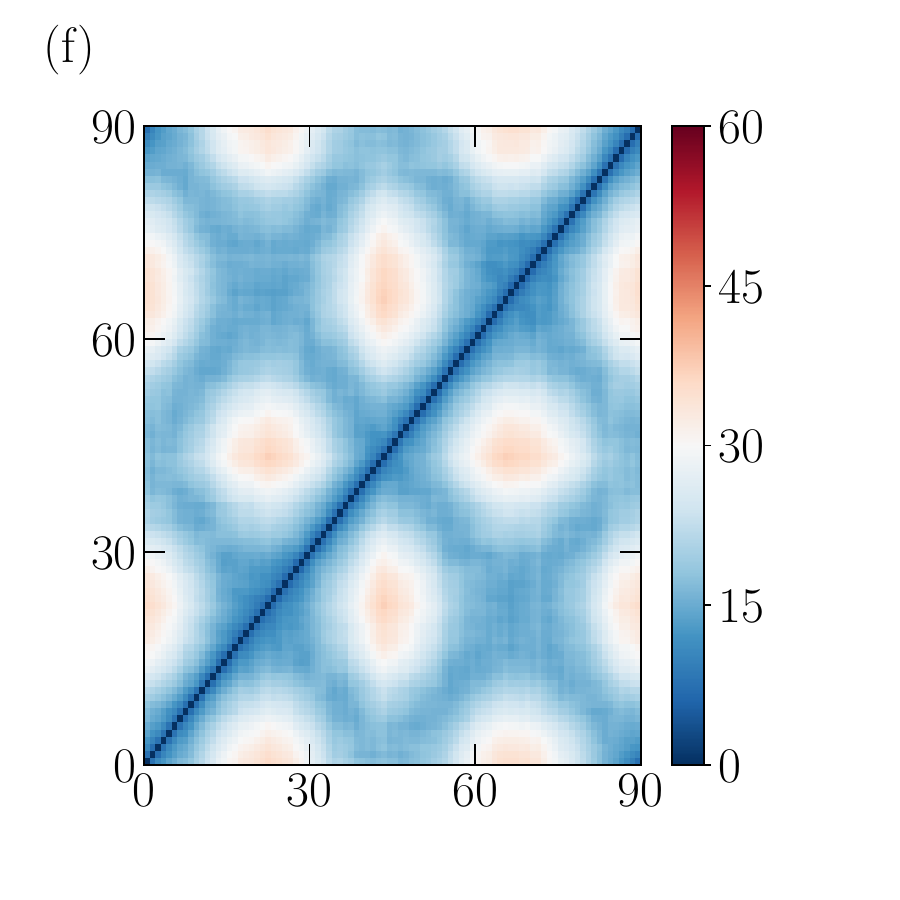}
    \caption{The Frobenius norm of the second-moment difference $\left \Vert \mb C(y) - \mb C(y')\right \Vert_F $ of the various CG models under the steady reverse Poiseuille flow generated by external force $f_0=0.01$ (upper) and $f_0=0.005$ (lower), where $y$ and $y'$ are represented by the horizontal and vertical axis, respectively. (a) and (d) CGCOM; (b) and (e) CG3; (c) and (f) CG4.}
    \label{fig:diff_second_momemt}
\end{figure}

Figure \ref{fig:diff_second_momemt} shows the difference $\left \Vert \mb C(y) - \mb C(y')\right \Vert_F $ of the various CG models at different locations, where $y$ and $y'$ correspond to the horizontal and vertical axis, respectively, and $\Vert \cdot \Vert_F$ is the Frobenius norm. It is clear that when the COMs are the only CG variables, the unresolved (i.e., orthogonal) degrees of freedom exhibit heterogeneous second moment along the $y\mhyphen$drection, implying pronounced variation of the conditional PDF under different external fields. On the other hand, the variation decreases significantly when auxiliary CG variables are introduced into the CG3 and CG4 models, showing the promise of certain generalization ability for non-equilibrium processes. 

The discrepancy in the conditional PDF will further lead to inconsistent reduced modeling terms under various non-equilibrium processes. To probe this effect, we examine the pairwise conservative force between the CG coordinates. 
We emphasize that the CG force generally exhibits a many-body nature \cite{Ge_Lei_JCP_2023}; here we use the projection along the pairwise direction \cite{Lei_Cas_2010, hijon2010mori} as a metric to quantify the generalization ability of the free energy $U(\mb Q)$.
Specifically, let $\mb F^{IJ}_{ij}$ denote the force between the CG coordinates $\mb Q^I_{i}$ and $\mb Q^J_{j}$.   
%
%
For the equilibrium state, the pairwise conservative force is defined as $F_{ij}^{\text{eq}}(Q) = \mathbb E^{\text{eq}}\left[\mb F^{IJ}_{ij}\cdot \mathbf e^{IJ}_{ij}\left|Q^{IJ}_{ij}=Q\right.\right]$, where
$\mathbf e^{IJ}_{ij} = \mb Q^{IJ}_{ij}/Q^{IJ}_{ij}$ and
$ \mathbb E^{\text{eq}}[\cdot]$ is the expectation over the molecule index $1\le I, J \le M$ under the equilibrium state, and $1\le i, j\le m$ is kept to represent the CG coordinate index within each molecule. For the non-equilibrium state, the pairwise conservative force is defined as $F_{ij}^{\text{neq}}(Q,y) = \mathbb{E}^{\text{neq}}\left[\mb{F}^{IJ}_{ij} \cdot \mathbf{e}^{IJ}_{ij} \mid y^I_i = y^J_j = y, Q^{IJ}_{ij} = Q \right]$, where the condition is taken that both $\mb{Q}_i^{I}$ and $\mb{Q}_j^{J}$ share a specific value along the $y\mhyphen$direction.

\begin{figure}
    \centering
    \includegraphics[width=0.3\textwidth]{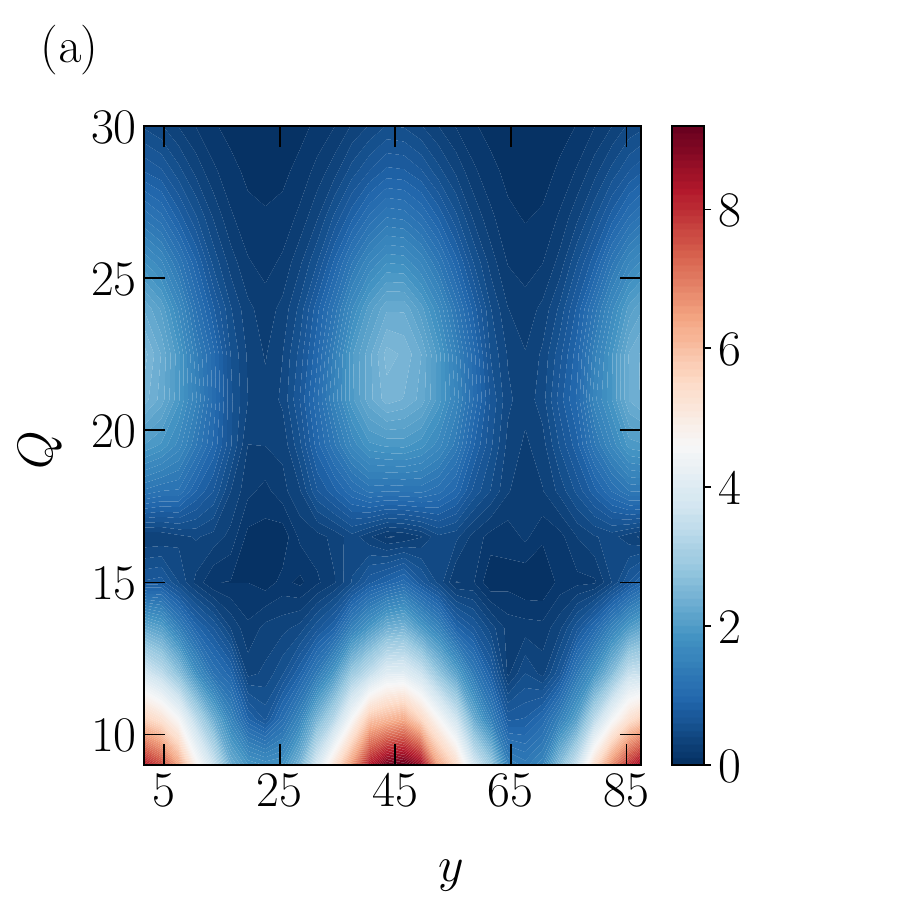}
    \includegraphics[width=0.3\textwidth]{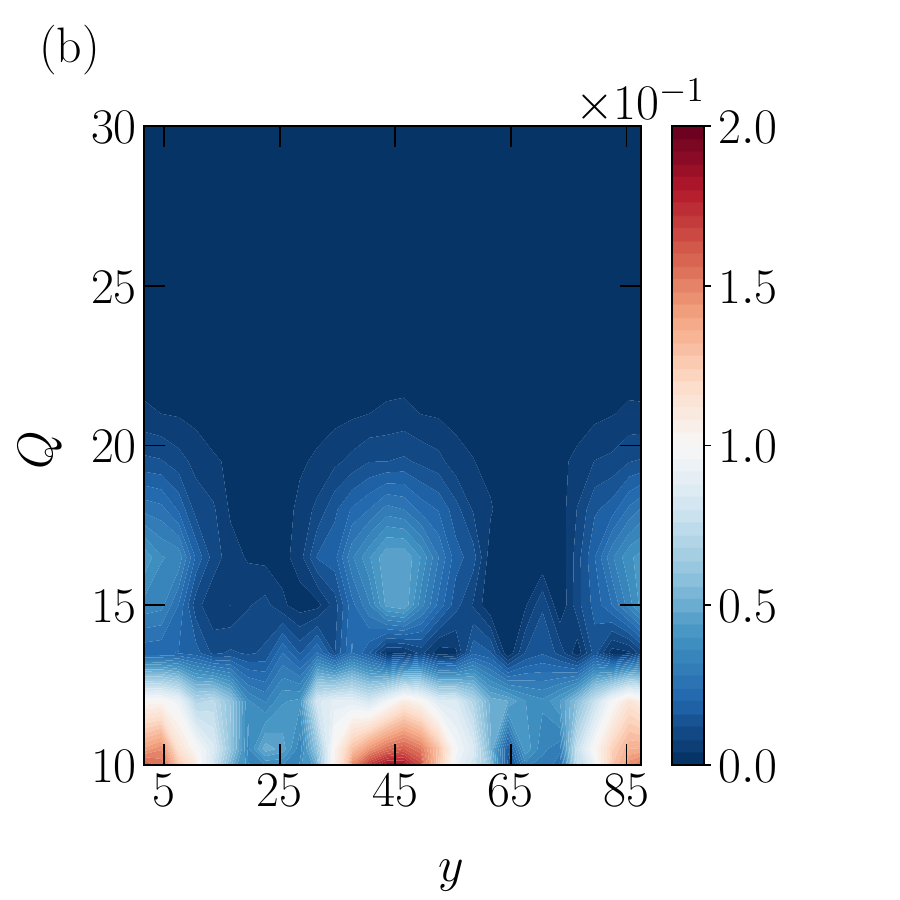}
    \includegraphics[width=0.3\textwidth]{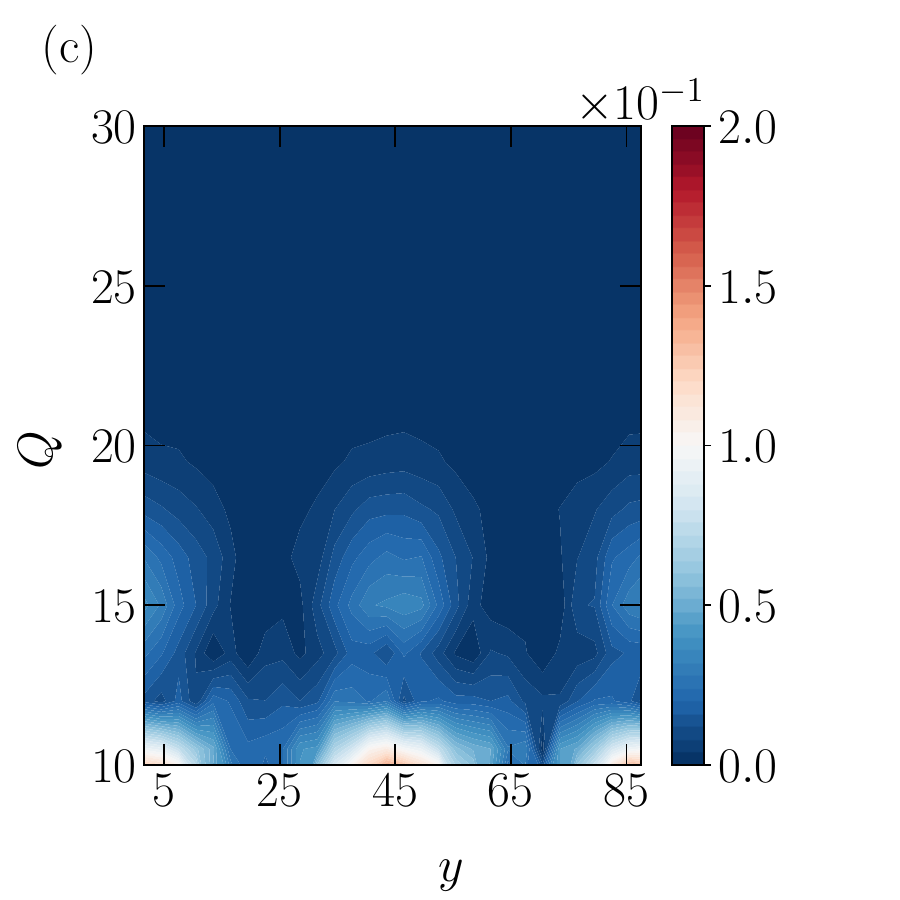}
    \includegraphics[width=0.3\textwidth]{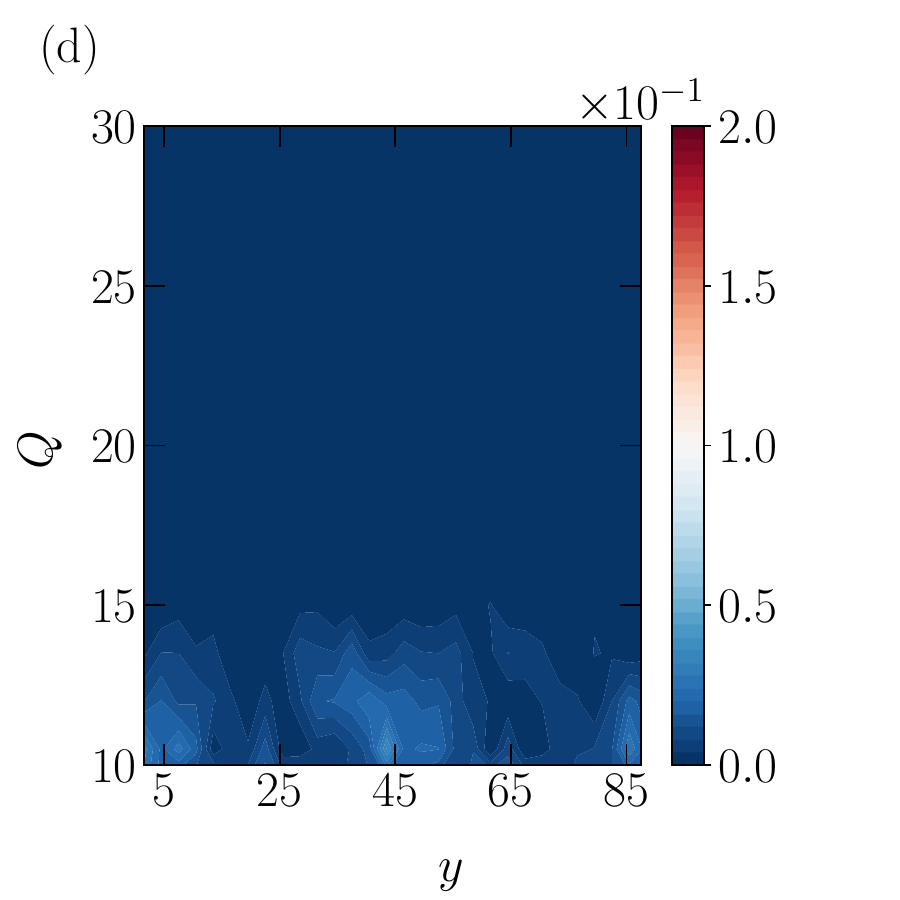}
    \includegraphics[width=0.3\textwidth]{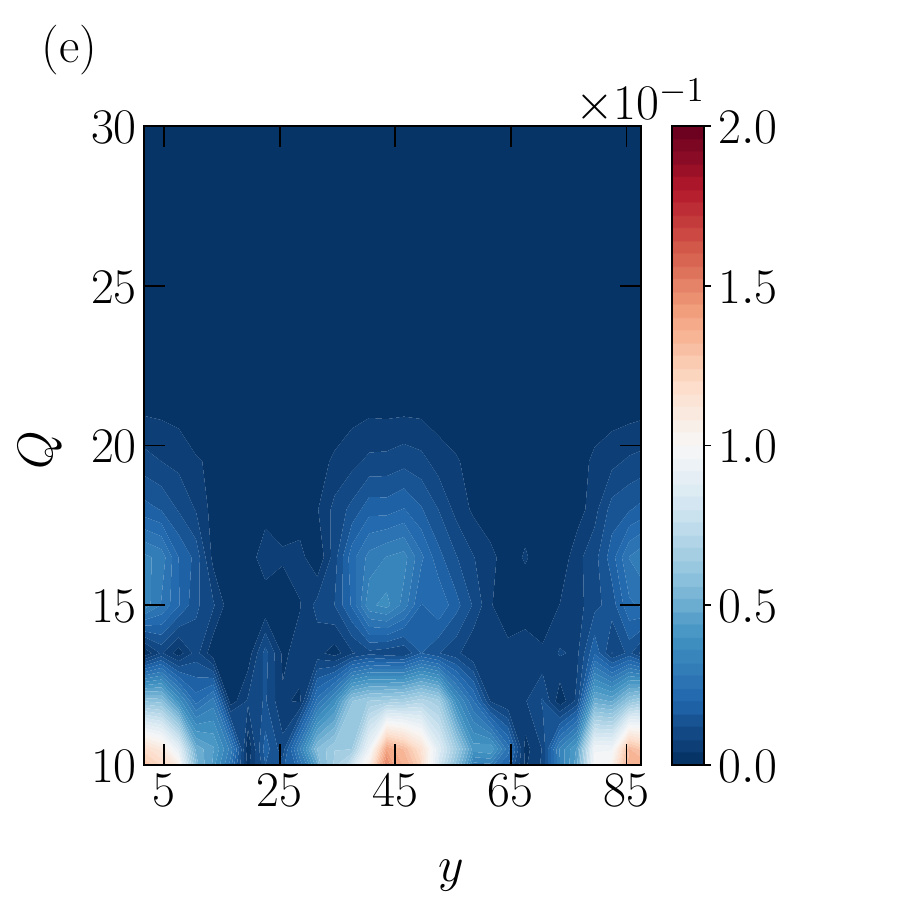}
    \includegraphics[width=0.3\textwidth]{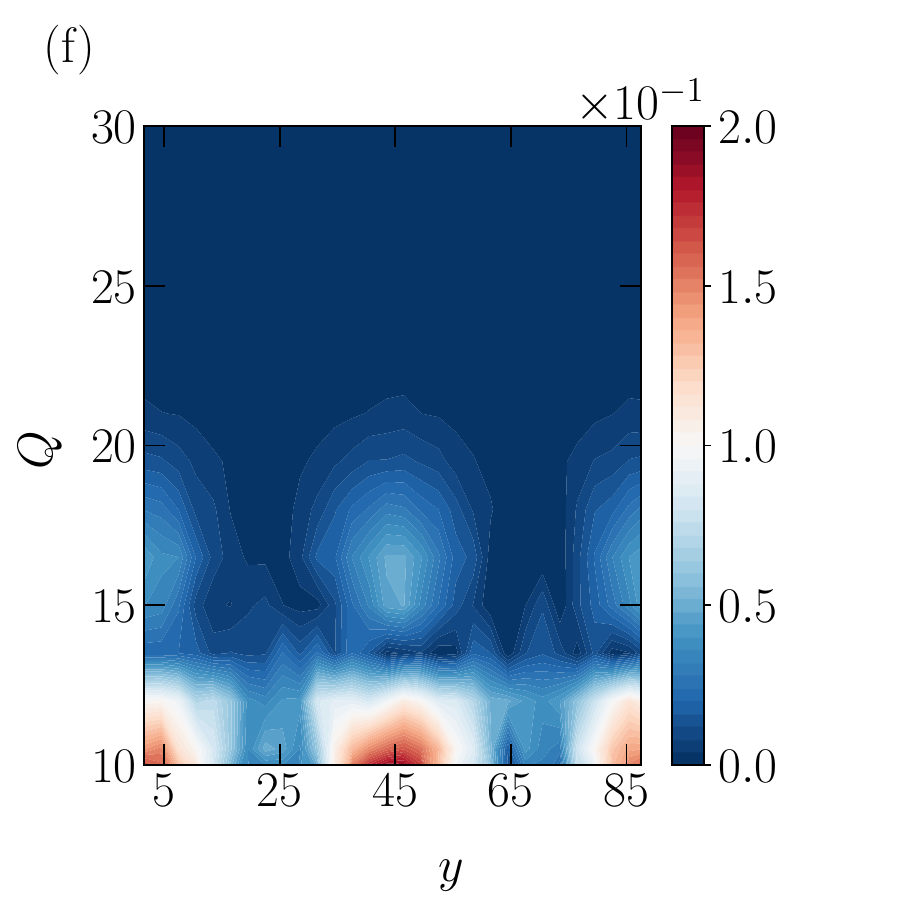}
    \caption{The pairwise conservative force difference $\left \vert F_{ij}^\text{neq}(Q) - F_{ij}^{\text{eq}}(Q,y)\right \vert$ for various CG models,  which loosely quantifies the generalization ability of the CG free energy $U(\mb Q)$. (a) CGCOM;  (b) CG3 with $(i,j) = (1,1)$; (c-f) CG4 with $(i,j) = (1,1), (2,2), \cdots, (4,4)$.}
    \label{fig:cf}
\end{figure}

Figure \ref{fig:cf} shows the difference 
$\left \vert F_{ij}^{\text{eq}}(Q) - F_{ij}^{\text{neq}}(Q,y)\right \vert $ for the different CG models. For the standard CGCOM model where $\mb Q$ is represented by the COMs, the two CG force terms agree well at $y=L/4 = 22.5$ and $y=3L/4= 67.5$, where the shear rate is close to zero and the atomistic particle distribution is near equilibrium. In contrast, the two force terms show pronounced differences at $y=0, 2/L$, and $L$ where the shear rate is large and the non-equilibrium distribution deviates from the equilibrium distribution. This is consistent with the specific pattern of the second-moment difference in Fig. \ref{fig:diff_second_momemt}(a) and (d). The large difference implies the heterogeneity of the conditional PDF $\delta(\phi^Q(\mb q) - \mb Q) \rho_e(\mb Q)$ under various external conditions. Such limitation has a clear physical interpretation: the intra-molecular interactions may significantly affect the visco-elastic response as well as the collective dynamics, which, however, cannot be captured by the COMs. Fortunately, the inconsistency can be alleviated by introducing auxiliary CG variables into the reduced model. As shown in Fig. \ref{fig:cf}, the conservative force difference $\vert F_{ij}^{\text{eq}}(Q) - F_{ij}^{\text{neq}}(Q,y)\vert $ decreases to $O(0.1)$ for the CG3 and CG4 models, showing the improvement in the generalization ability of the free energy term $U(\mb Q)$.  

\begin{figure}
    \centering
    \includegraphics[width=0.3\textwidth]{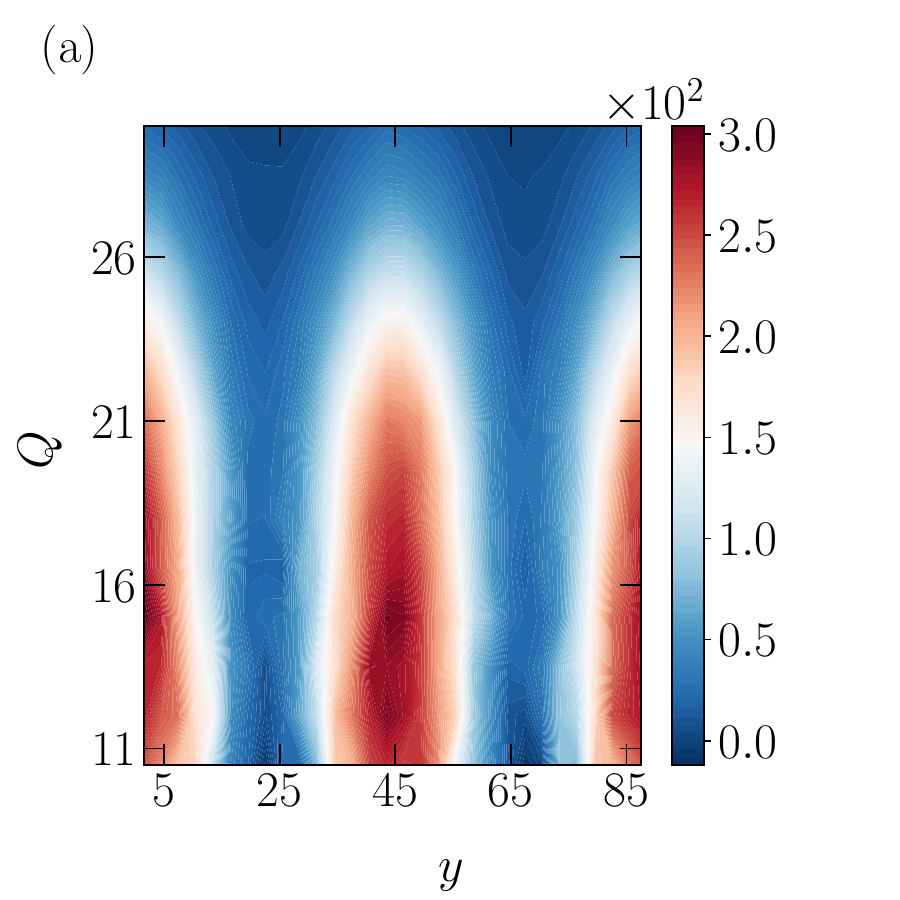}
    \includegraphics[width=0.3\textwidth]{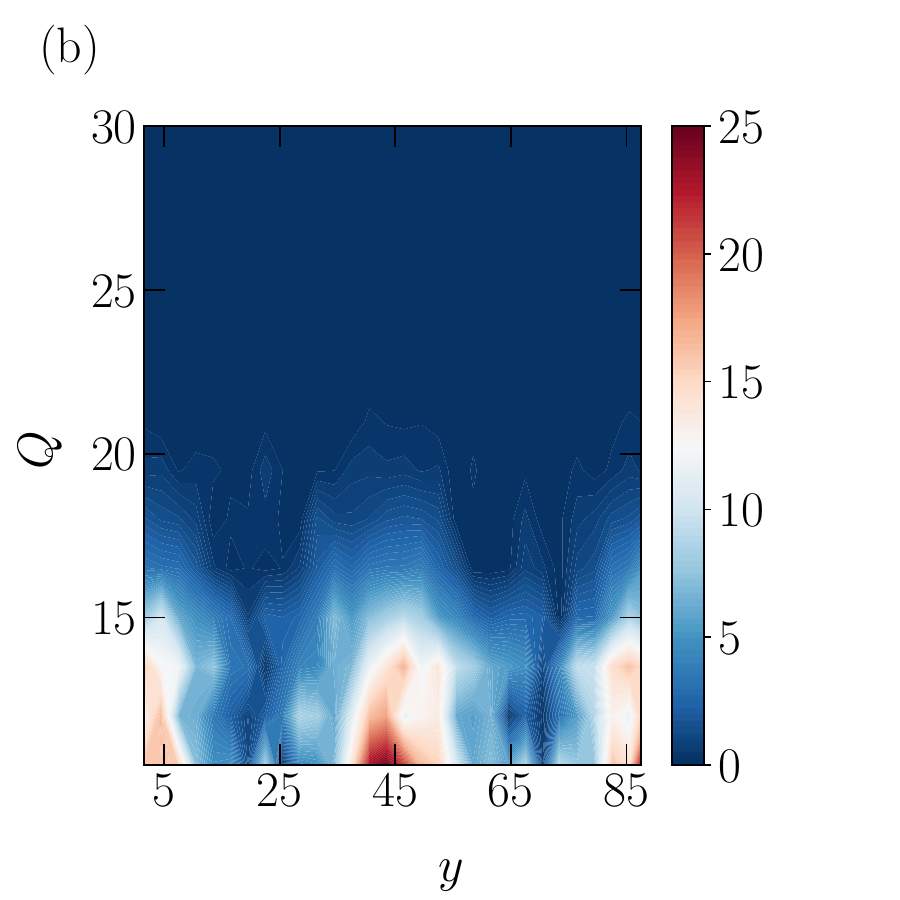}
    \includegraphics[width=0.3\textwidth]{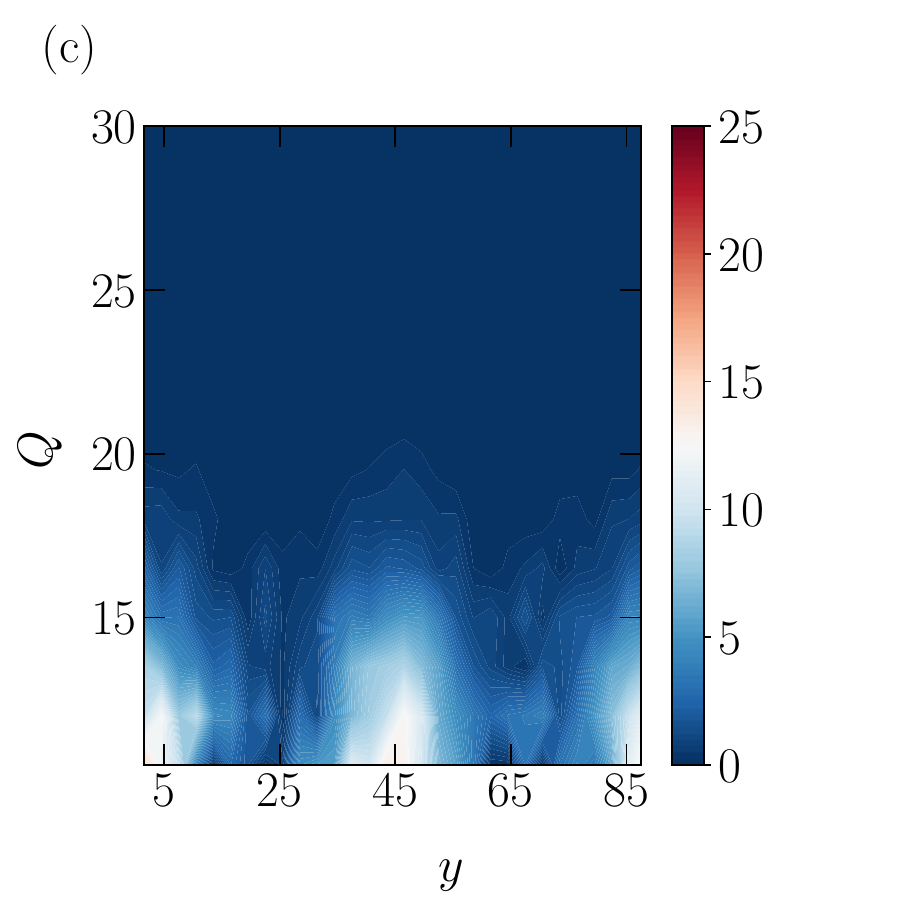}
    \includegraphics[width=0.3\textwidth]{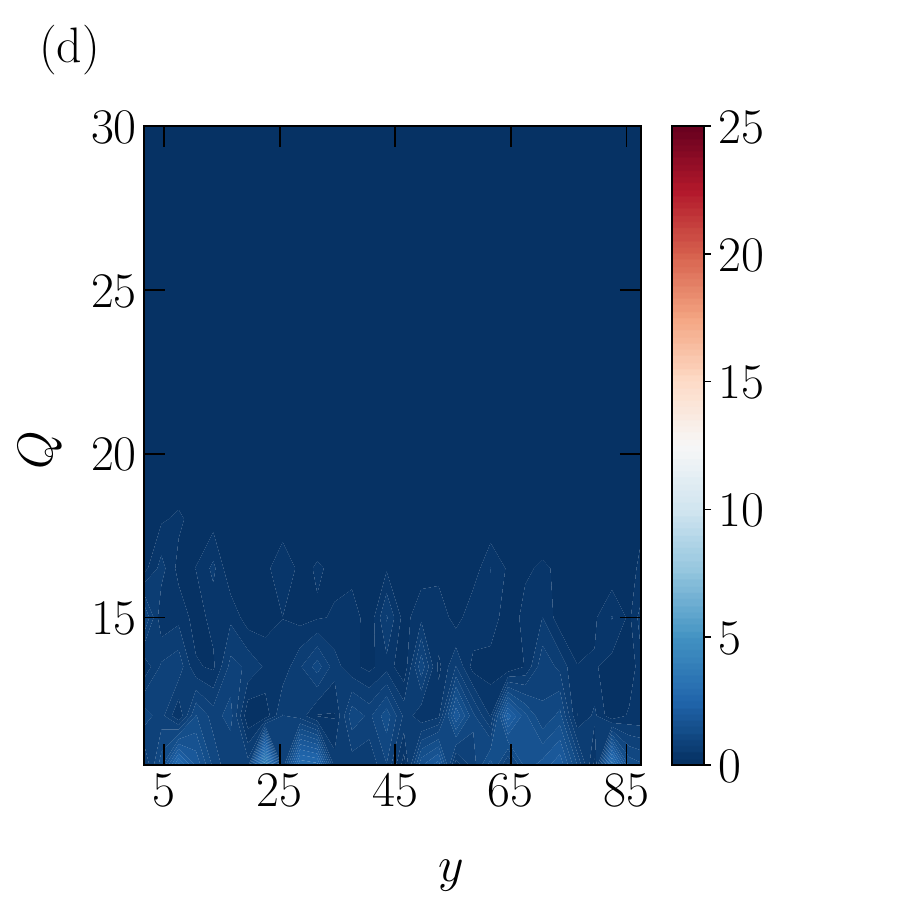}
    \includegraphics[width=0.3\textwidth]{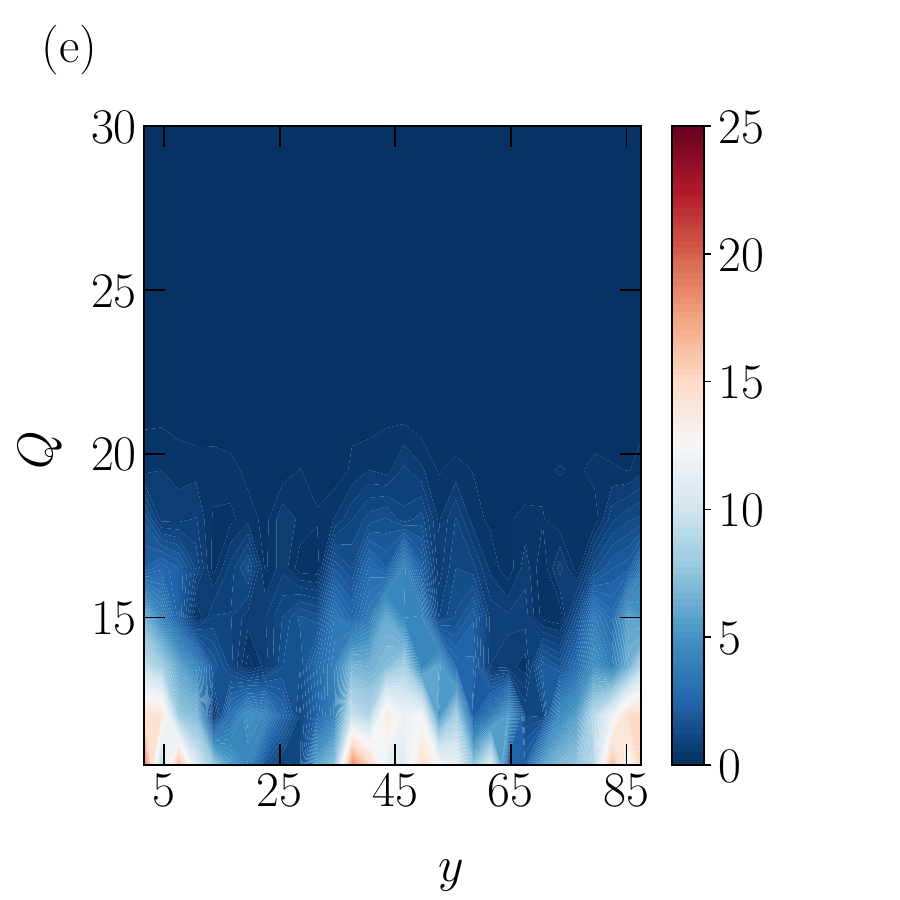}
    \includegraphics[width=0.3\textwidth]{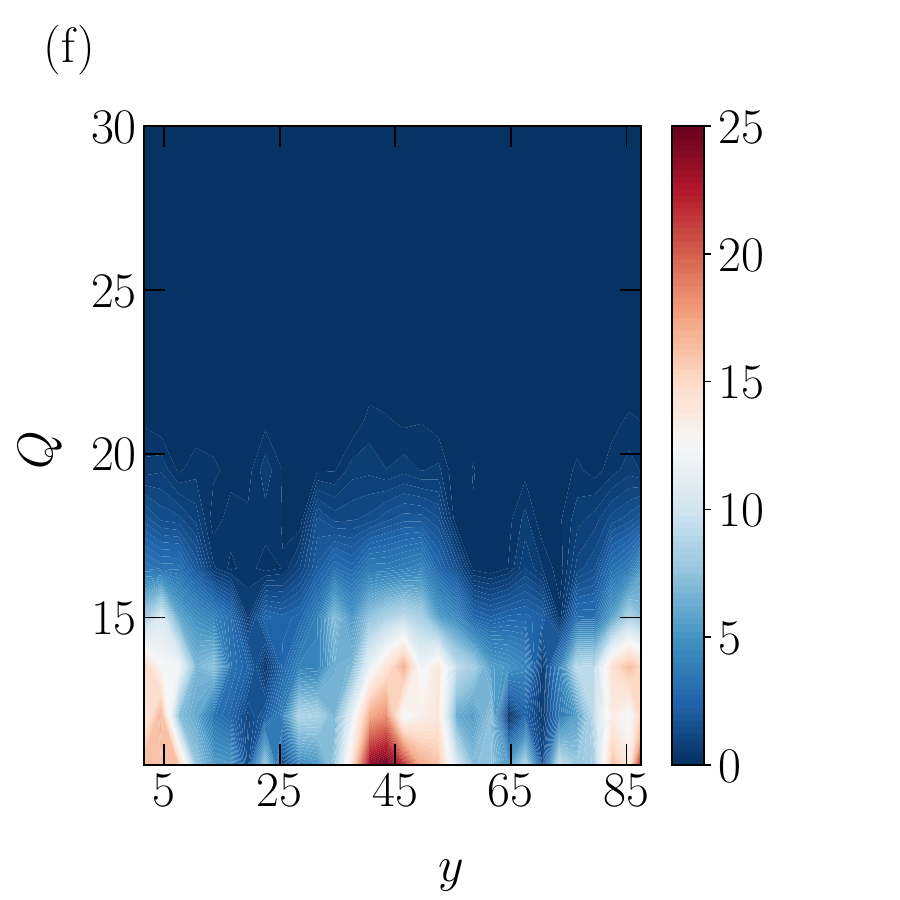}
    \caption{The difference of the variation of the fluctuation force $\vert K^{ij}_\text{neq}(Q)- K^{ij}_{\text{eq}}(Q,y)\vert $ for different CG models, which loosely quantifies the generalization ability of the CG memory term $\mb K(\mb Q, t)$ at $t=0$. The sub-figure labels are the same as Fig. \ref{fig:cf}. }
    \label{fig:mk}
\end{figure}

Finally, we study the generalization ability of the memory term $\mb K(\mb Q, t)$. Similar to $\mb U(\mb Q)$, we note that the memory term $\mb K(\mb Q, t)$ generally exhibits a many-body nature \cite{lyu2023construction}; here we examine the variation of the pairwise fluctuation force $\delta \mb F^{IJ}_{ij} =  \mb F^{IJ}_{ij} - F_{ij}(Q^{IJ}_{ij}) \mathbf e^{IJ}_{ij}$. It can be loosely viewed as a metric of $\mb K(\mb Q, t=0) = \mathbb{E} \left[\delta \mb F \otimes \delta\mb F\right]$. Specifically, we evaluate the variation of $\delta \mb F^{IJ}_{ij}$
under the equilibrium state  
$$
K_{ij}^{\text{eq}}(Q) = \mathbb E^{\text{eq}}\left[{\delta \mb F^{IJ}_{ij}}^T \left(\mb I - {\mb e^{IJ}_{ij} }^T\mb e^{IJ}_{ij}\right ) \delta \mb F^{IJ}_{ij} \left|Q^{IJ}_{ij}=Q \right.\right],
$$ and the nonequilibrium state
$$
K_{ij}^{\text{neq}}(Q,y) = \mathbb E^{\text{neq}}\left[{\delta \mb F^{IJ}_{ij}}^T \left(\mb I - {\mb e^{IJ}_{ij} }^T\mb e^{IJ}_{ij}\right ) \delta \mb F^{IJ}_{ij} \left|Q^{IJ}_{ij}=Q ,y^I_i =y^J_j = y\right.\right],
$$
where the expectation is over the molecule index $1\le I, J \le M$. Fig. \ref{fig:mk} shows the difference $\vert K_{ij}^{\rm neq}(Q, y) - K_{ij}^{\rm eq}(Q)\vert$ for various CG models. Similar to the conservative CG force, the variation of the fluctuation force exhibits pronounced heterogeneity for the CGCOM model and achieves the largest discrepancy at the locations of large shear rate ($y = 0, L/2, L$). 
In contrast, the discrepancy becomes much smaller for the CG3 and CG4 models with auxiliary CG variables. 

\subsection{Non-equilibrium flows}
The inconsistent reduced modeling terms shown in Sec. \ref{sec:generalization_CG} reveal a caveat of using the standard CGCOM model for non-equilibrium processes. We examine this effect by simulating various non-equilibrium flows. 

To establish a fair comparison, we construct the many-body form of the free energy $U(\mb Q)$ and the memory term $\mb K(\mb Q, t)$ with the symmetry-preserving neural network representations presented in Sec. \ref{sec:free_energy} and Sec. \ref{sec:memory} for the CGCOM, CG3 and CG4 models, respectively. To verify $U(\mb Q)$, we sample the radial distribution function of the inter-molecular COMs for the CGCOM model and the intra-molecular CG coordinates for the CG3 and CG4 models. 
To verify $\mb K(\mb Q,t)$, we sample the velocity auto-correlation function of the CG variables. As shown in the Appendix, the predictions of the three CG models show good agreement with the full MD results, which verify their efficacy for modeling equilibrium processes.

\begin{figure}
    \centering
    \includegraphics[width=0.45\textwidth]{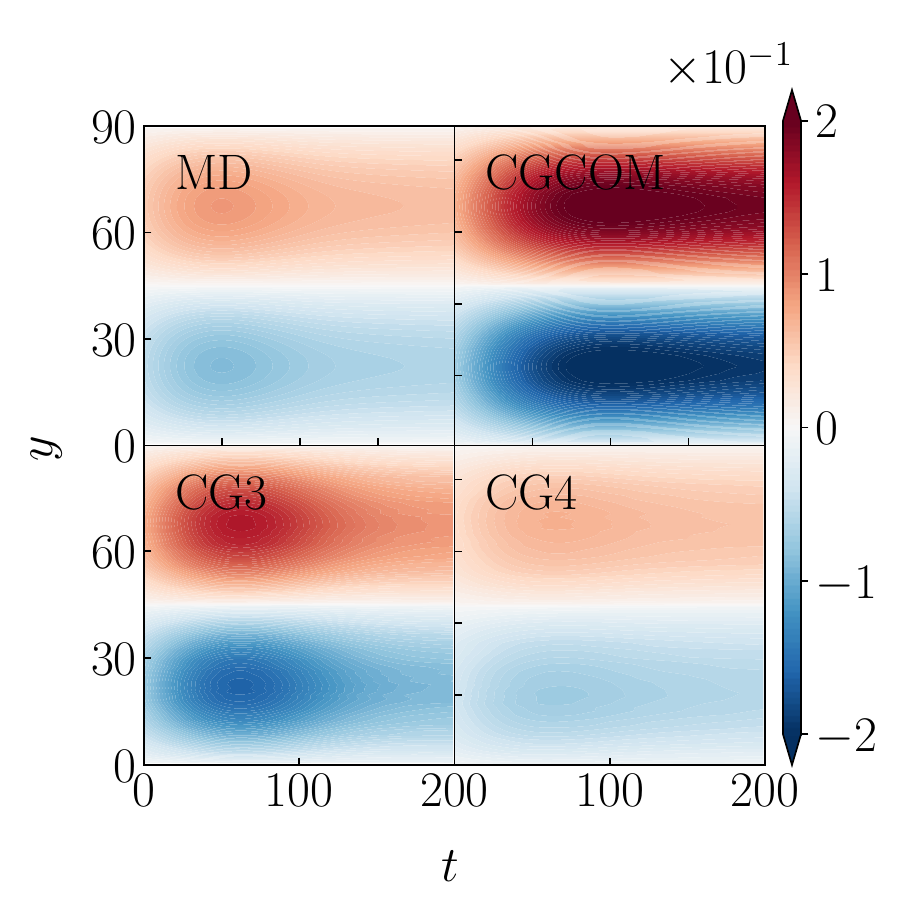}
    \caption{The development of the reverse Poiseuille flow under the external force $f_0 = 0.01$ predicted by the full MD and three CG models.}
    \label{fig:noneq_dev}
\end{figure}

Next, we consider the non-equilibrium reverse Poiseuille flow generated by an external force \eqref{eq:external_force} with $f_0 = 0.01$. Fig. \ref{fig:noneq_dev} shows the velocity development $u_x(y, t)$. Compared with the full MD results, the prediction of the standard CGCOM overestimates the instantaneous velocity magnitude by three times. Furthermore, it can not capture the development oscillation near $t = 60$. This limitation is rooted in the choice of the COMs as the CG variables, which ignores the intra-molecular interactions and therefore can not capture the complex visco-elastic responses under non-equilibrium processes.  The prediction of the CG3 model shows significant improvement to the CGCOM model but overestimates the magnitude of the velocity oscillation during the flow development stage. On the other hand, the prediction of the CG4 model shows a good agreement with the full MD results throughout the development stage.

\begin{figure}
    \centering
    \includegraphics[width=0.45\textwidth]{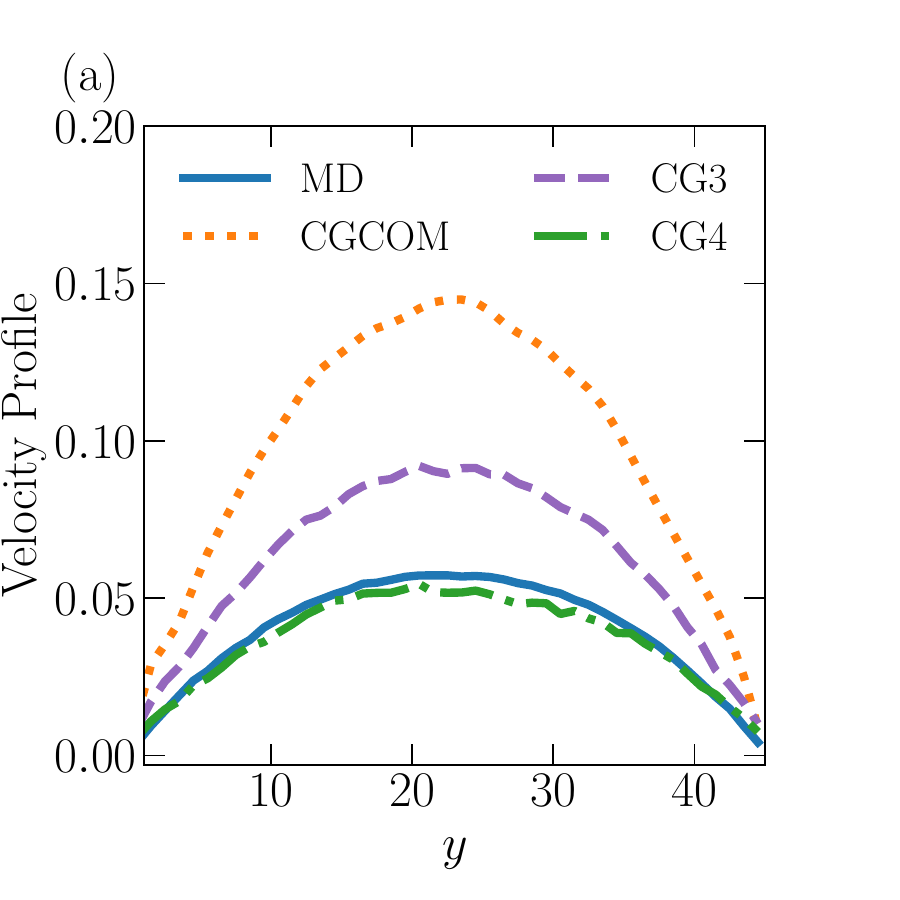}
    \includegraphics[width=0.45\textwidth]{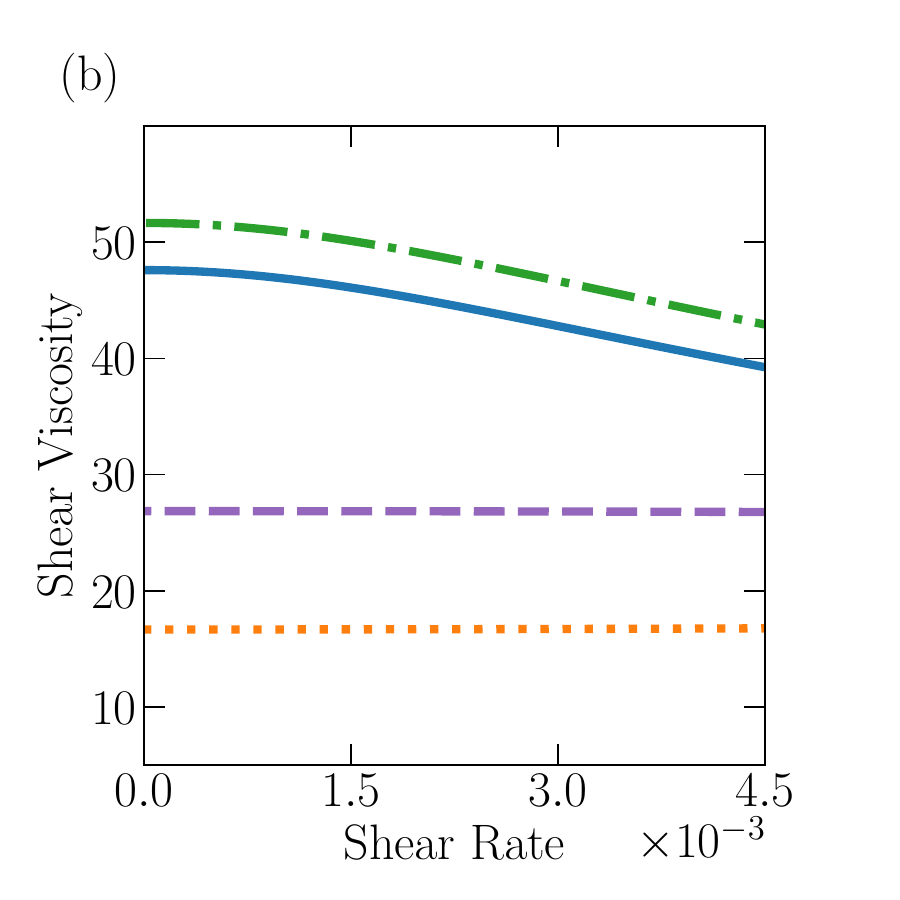}
    \caption{The steady state reverse Poiseuille flow under external force $f_0 = 0.01$ predicted by the full MD and the CG models. (a) The stationary velocity profile $u_x(y)$ (b) The shear-rate-dependent viscosity.}
    \label{fig:noneq}
\end{figure}

Fig. \ref{fig:noneq} shows the steady state velocity profile $u_x(y)$ predicted by the different models. Similar to the development stage, the prediction of the CG4 model shows a good agreement with the MD results. However, the prediction of the CGCOM and the CG3 models show apparent discrepancies due to their inefficacy in modeling the intra-molecular interactions. To quantify the difference, we compute the shear-rate-dependent viscosity from the steady velocity profile. The predictions of the CGCOM and the CG3 model are nearly independent of the shear rate, which indicates that they can not capture the visco-elastic responses associated with the molecular conformation change under external shear flow. In contrast, the CG4 model can faithfully reproduce the 
shear-rate-dependent viscosity of the full MD results.

\begin{figure}
    \centering
    \includegraphics[width=0.45\textwidth]{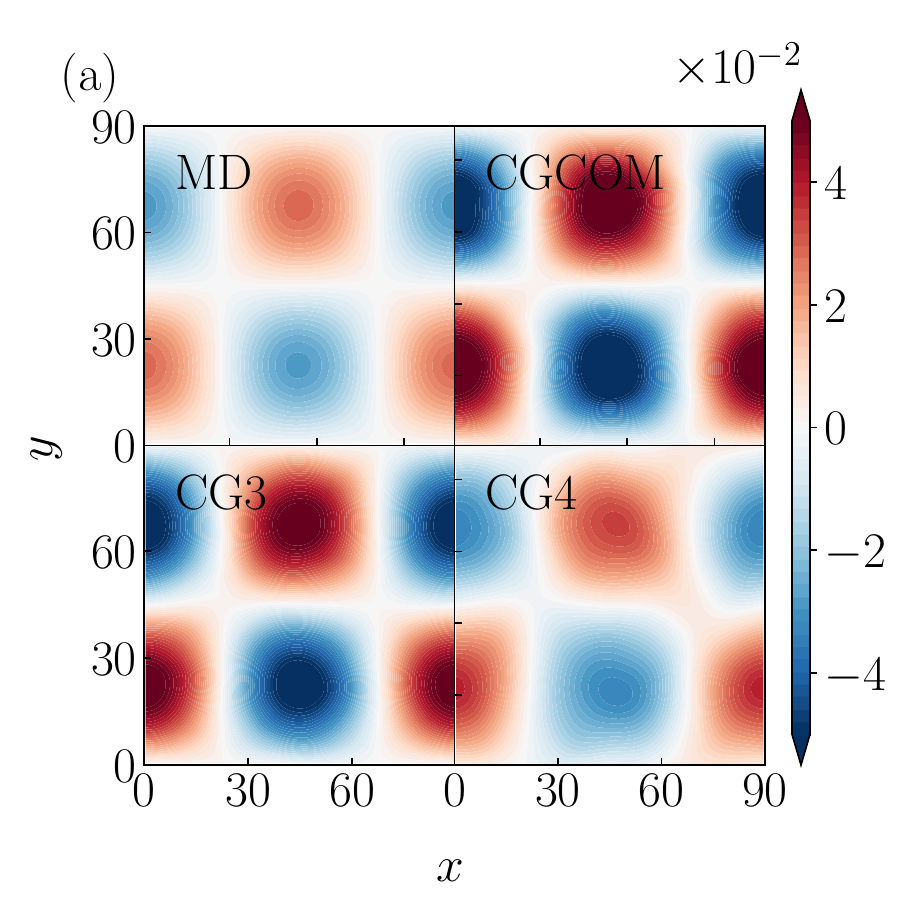}
    \includegraphics[width=0.45\textwidth]{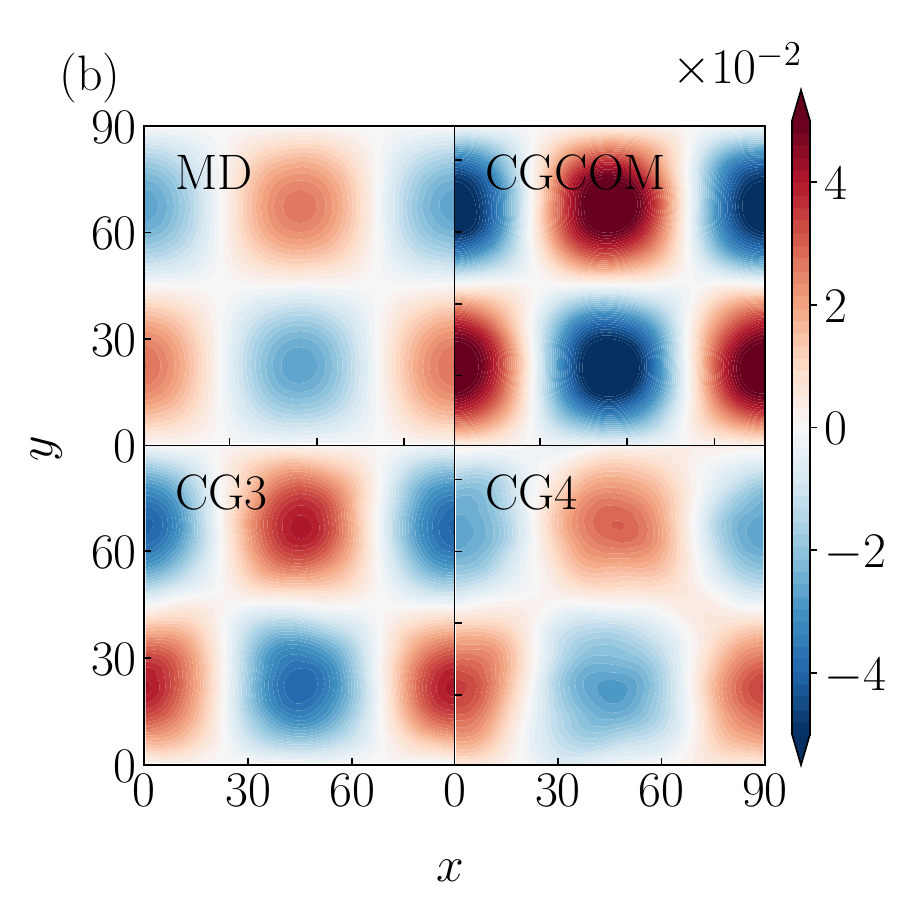}
      \includegraphics[width=0.45\textwidth]{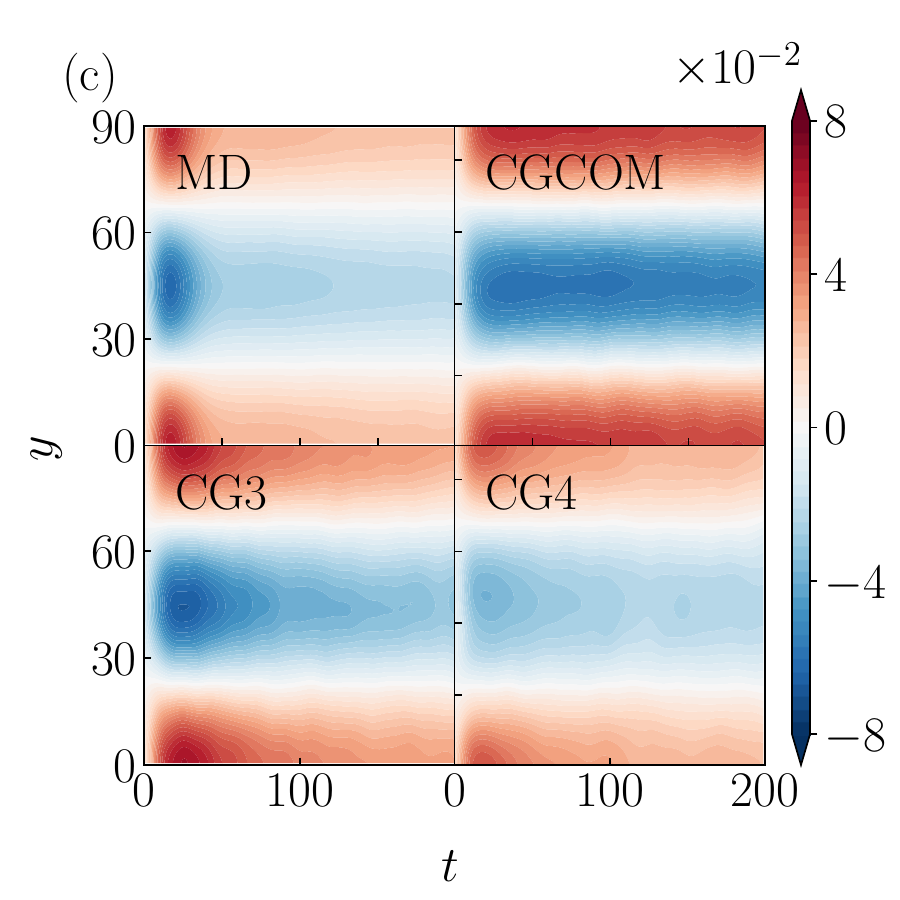}
    \includegraphics[width=0.45\textwidth]{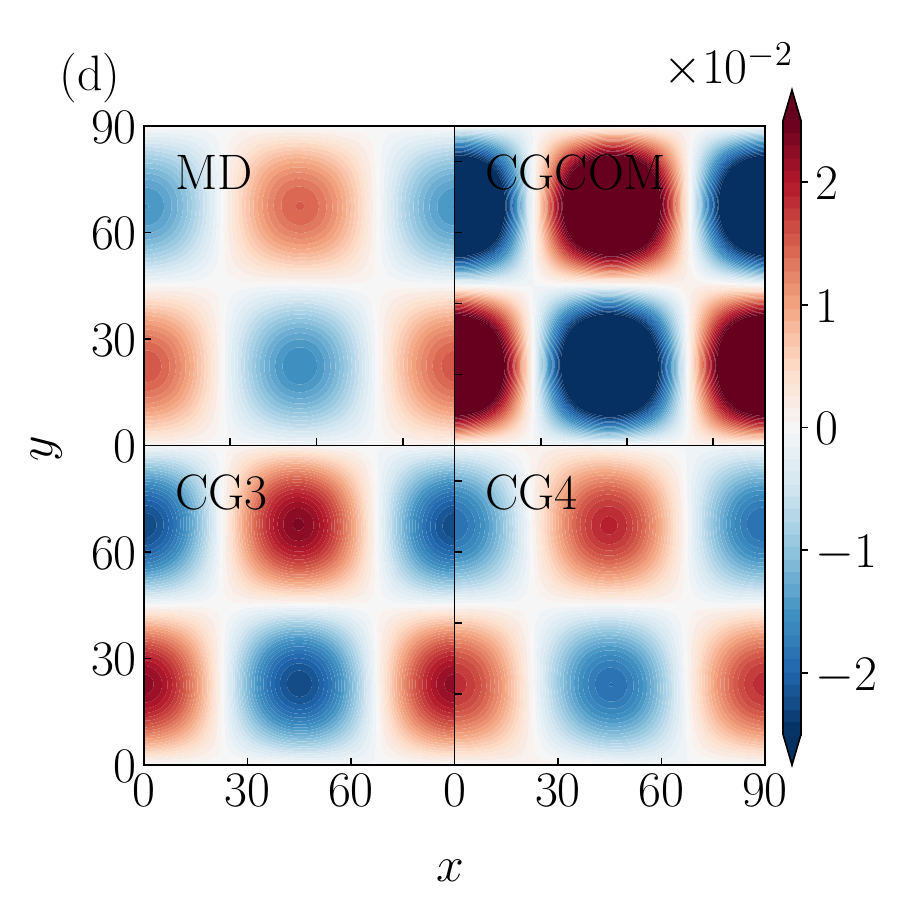}
    \caption{The velocity field of a 2D Taylor vortex predicted by the full MD and the CG models. (a-b) The contour of the instantaneous velocity $u_y(x,y,t)$ at $t=50$ (a) and $t=100$ (b). (c) The development of velocity $u_y(x=25, y, t)$ (d) The steady velocity field $u_y(x,y)$.}
    \label{fig:vortex}
\end{figure}

Finally, we examine the generalization ability of the CG models for other flow fields. While the training samples of the CG models are collected in the reverse Poiseuille flow, we validate the constructed models with the Taylor vortex generated by the external force field  
\begin{equation*} 
f_x(x,y) = -f_0 \sin\left(\frac{2\pi x}{L}\right)\cos\left(\frac{2\pi y}{L}\right), \quad f_y(x,y) = f_0 \cos\left(\frac{2\pi x}{L}\right)\sin\left(\frac{2\pi y}{L}\right), 
\end{equation*} 
where $f_0=1\times 10^{-2}$ and $L=90$. Fig. \ref{fig:vortex} shows the prediction of the 2D velocity contour from various models. Similar to the previous example, the CGCOM and CG3 models overestimate the magnitude of the velocity field due to their insufficiency in modeling the intra-molecular interactions arising from the molecular conformation change under the external flow field. In contrast, the prediction of the CG4 model yields show good agreement with the full MD model.

Although the CG4 model performs well in the vortex flow, we should not view it as the ``correct'' (as opposed to the CGCOM) model for all non-equilibrium processes. Our main point is that the conditional PDF associated with the CG projection needs to be consistent to ensure the model's generalization ability, which, unfortunately, can not be guaranteed for non-equilibrium processes. This issue may severely limit a CG model's applicability to practical problems. On the other hand, properly introducing auxiliary CG variables may mitigate this inconsistency, and significantly improve the model's generalization ability.  

\section{Summary}
This work presents a caveat in constructing reliable coarse-grained molecular dynamics models for non-equilibrium processes. Specifically, a CG model's generalization ability relies on its consistency in the 
conditional PDF of the phase space vector (or equivalently, the CG projector operator) under various non-equilibrium conditions. This criterion is determined by the proper choice of the CG variables \emph{a priori} and can not be remedied by constructing more accurate reduced modeling (the free energy and the memory) terms.  
Although the Zwanzig's projection formalism can in principle provide the rigorous reduced dynamics, it is valid only when the distribution of the CG variables is consistent with the one associated with the CG projection operator. 
Unfortunately, this metric seems to be broadly overlooked in most existing studies on CGMD modeling. 
While the previous works (e.g. Refs. \cite{Lei_Cas_2010, hijon2010mori}) use dynamic properties such as the velocity auto-correlation functions and the mean squared displacement to validate the CG models, these quantities are essentially defined under the marginal density distribution near-equilibrium; their applications to non-equilibrium processes are generally un-warranted.

To alleviate the above challenge, this work proposes to learn a set of auxiliary CG variables based on the TICA to minimize the entropy contribution of the unresolved variables so that the conditional PDF of the full phase space vector under various conditions approaches the equilibrium case (i.e., Eq.  \eqref{eq:conditional_prob_metric}). We verify this generalization metric by examining the distribution of the second moment, the fluctuation force, and its variation under various external conditions. We show that the common CG model that uses the molecule's COMs as the CG variables is generally insufficient to retain the consistent conditional PDF. In contrast, the present models with auxiliary CG variables show significant improvement. Furthermore, the crucial role of this metric is reflected in modeling the non-equilibrium reverse Poiseuille flow and the Taylor vortex. In particular, the prediction of the standard CGCOM model exhibits large discrepancies from the full MD results due to its inefficacy in modeling the intramolecular interactions under external flows different from the equilibrium conditions. Conversely, the present model can faithfully recover the complex visco-elastic responses and therefore yields consistent predictions 
with the full MD results. 

Finally, we note that the learning of the auxiliary CG variables in the present study remains somewhat empirical. They are constructed based on the TICA and hence take a linear form of the full atomistic coordinates. In practice, we may learn the CG variables as nonlinear encoders of the molecular conformation to achieve a more efficient representation of the intra-molecular interactions (e.g., see Ref. \cite{Lei_E_DeePN2_2022}). We leave this for future study.

\appendix
\section{Additional result of CGMD model under equilibrium state}
\begin{figure}
    \centering
    \includegraphics[width=0.3\textwidth]{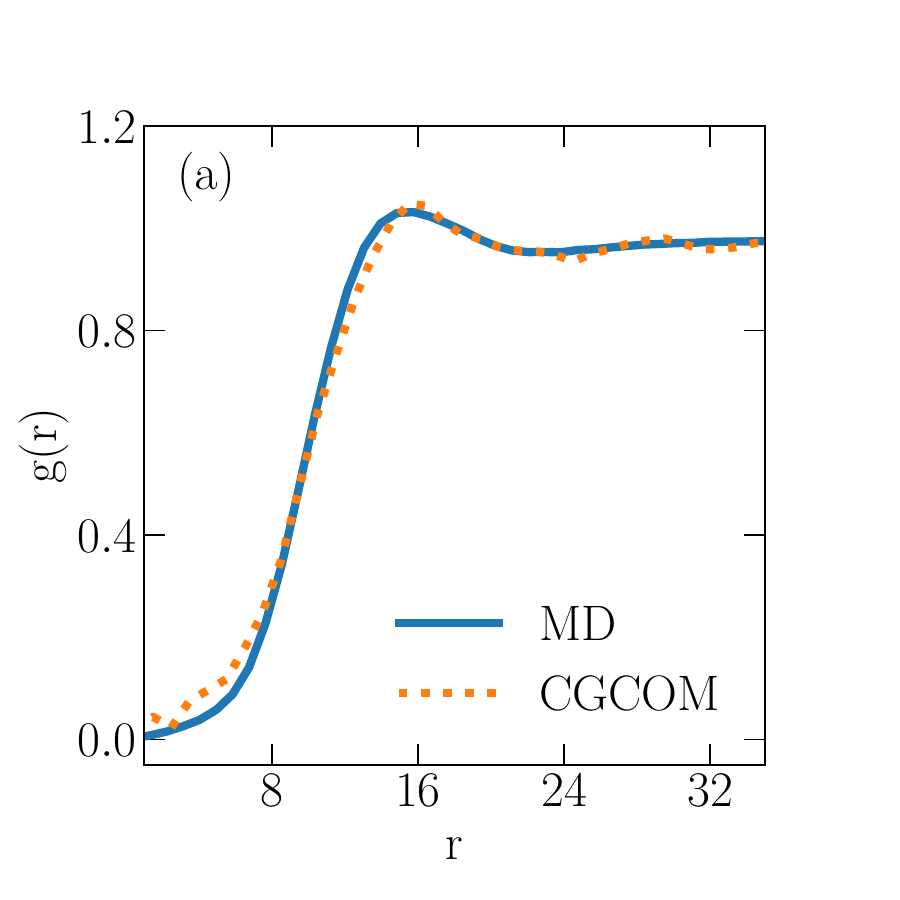}
    \includegraphics[width=0.3\textwidth]{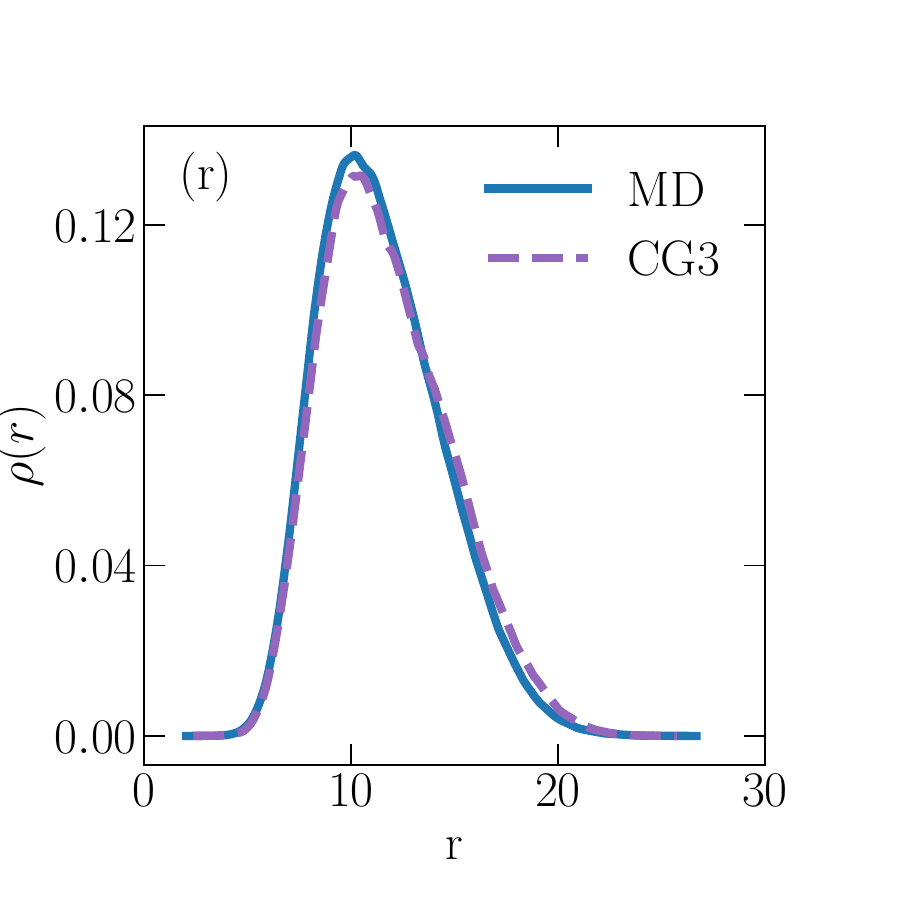}
    \includegraphics[width=0.3\textwidth]{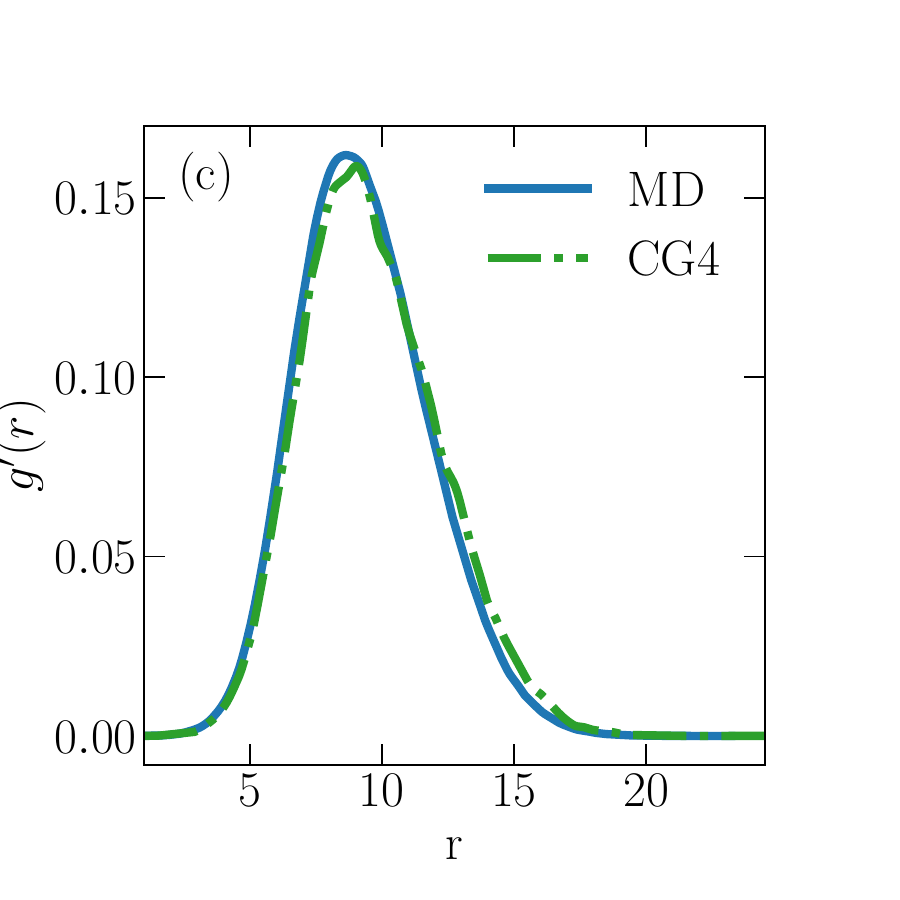}
    \includegraphics[width=0.3\textwidth]{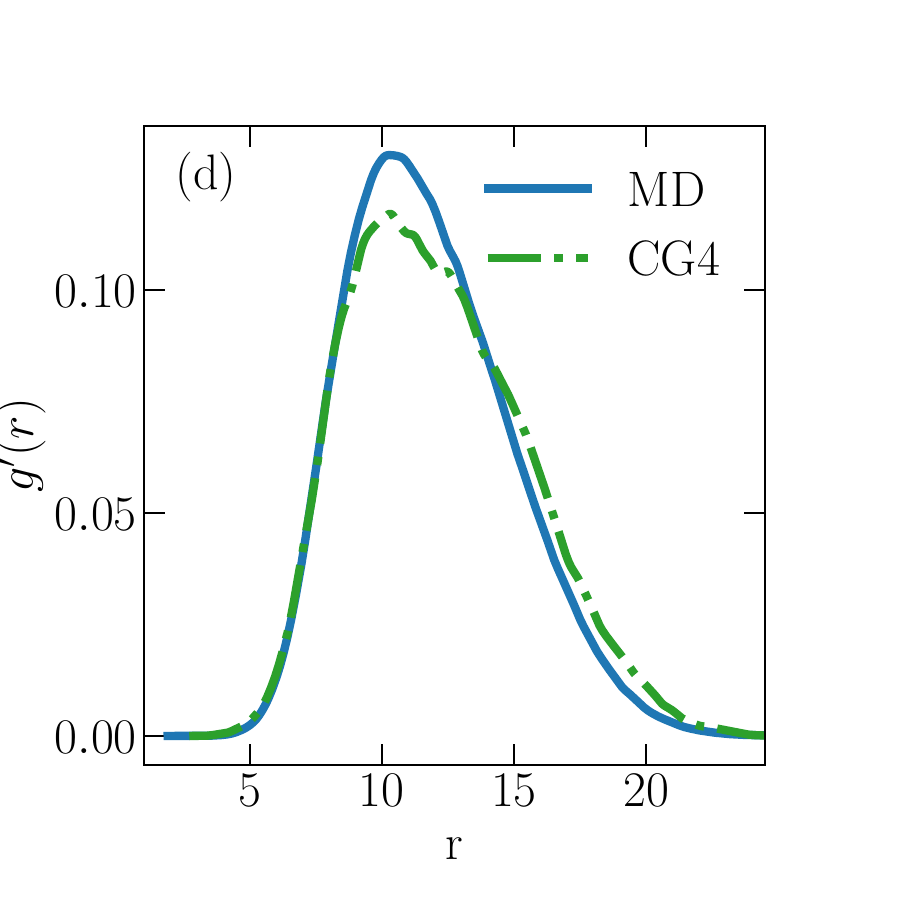}
    \includegraphics[width=0.3\textwidth]{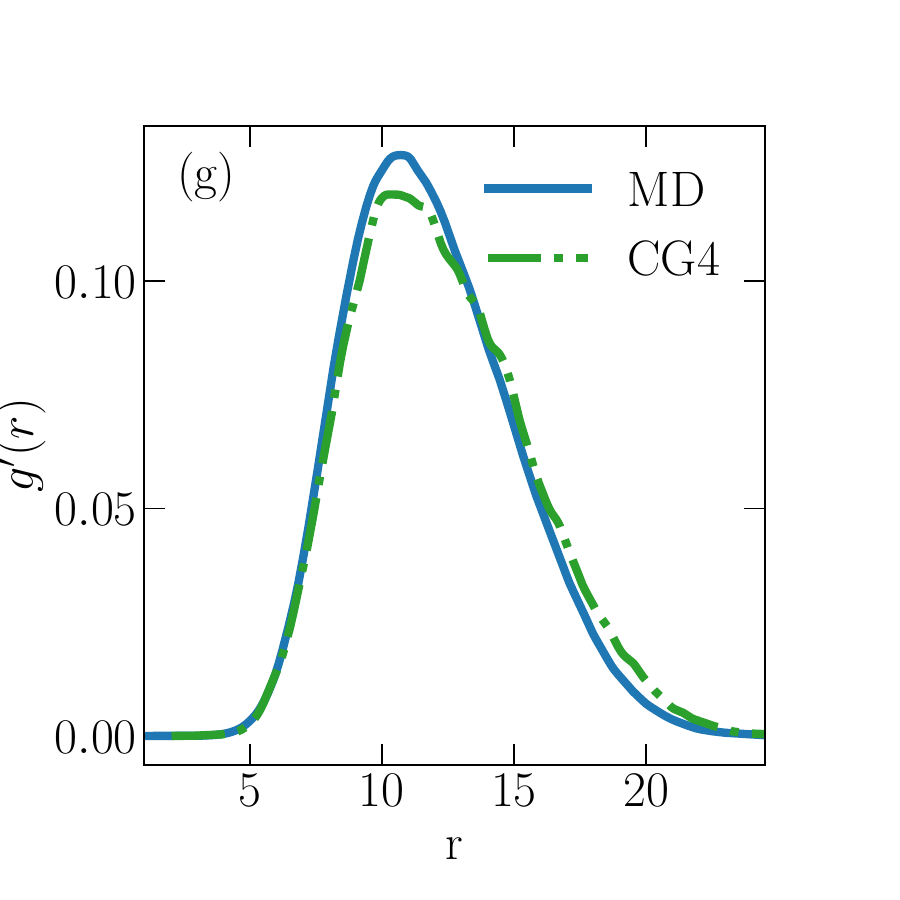}
    \includegraphics[width=0.3\textwidth]{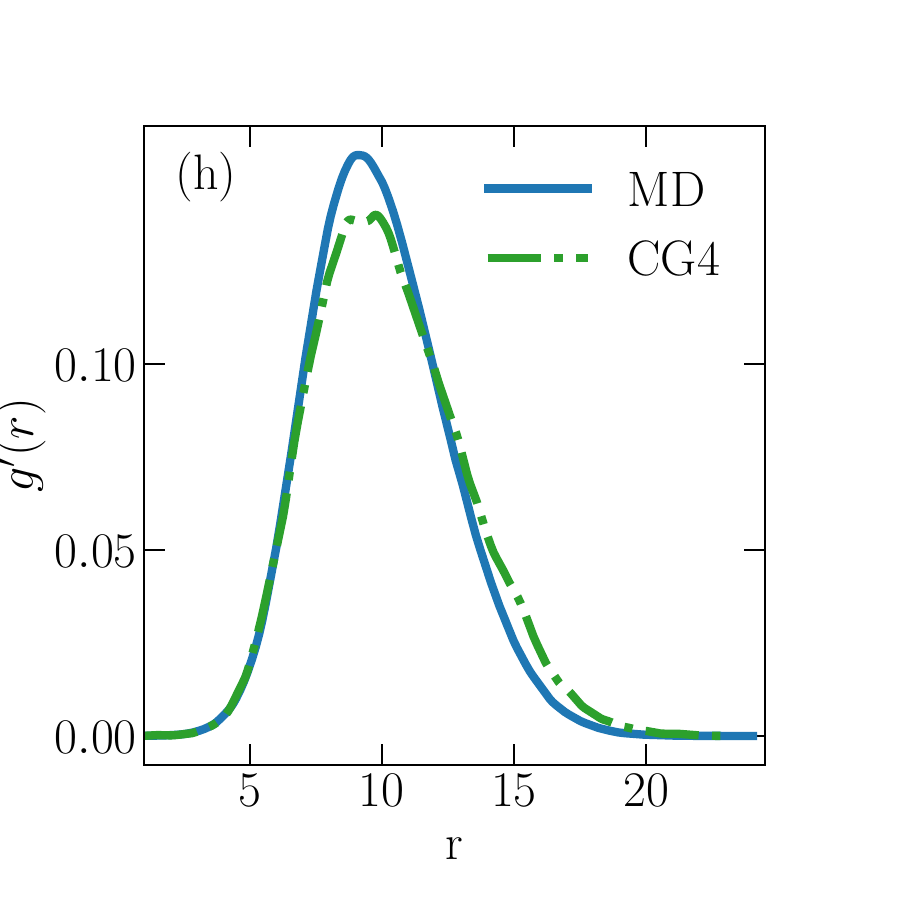}
    \includegraphics[width=0.3\textwidth]{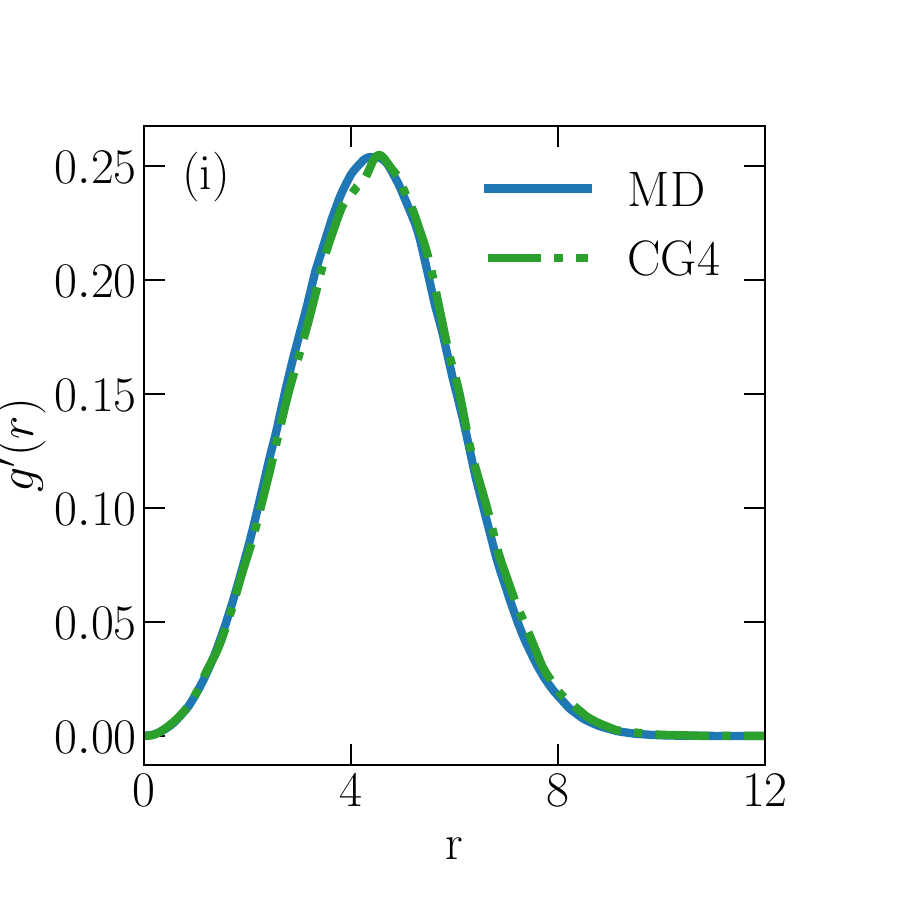}
    \includegraphics[width=0.3\textwidth]{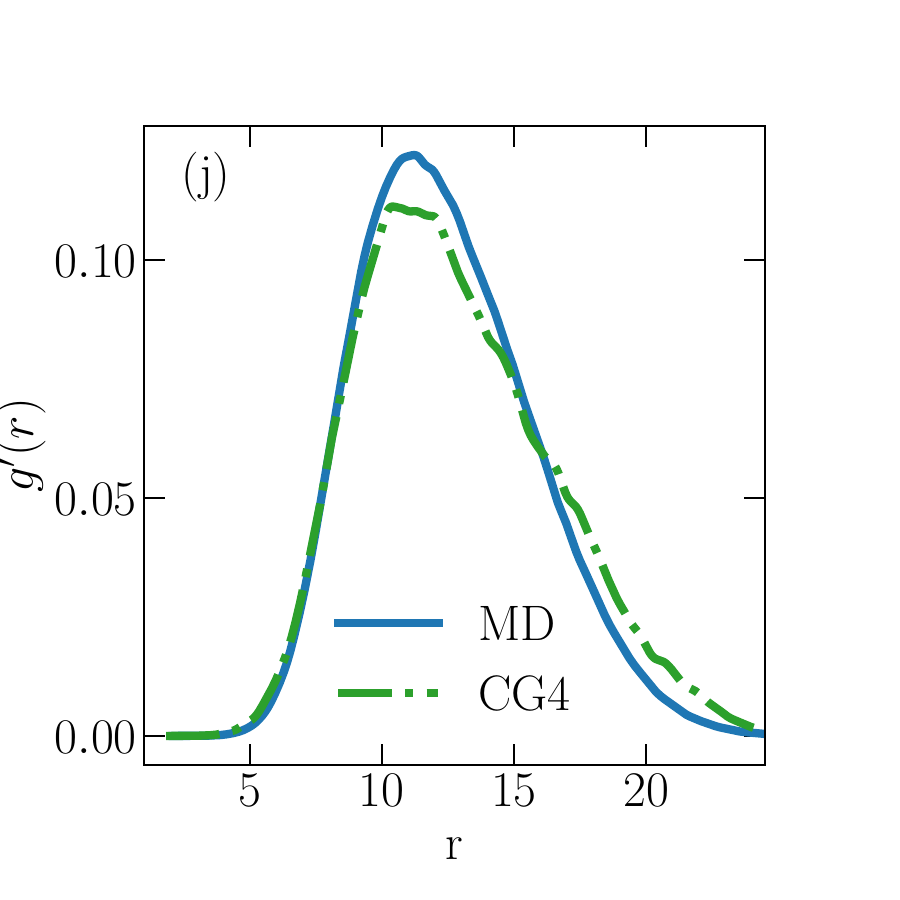}
    \caption{The distribution of the pairwise distance between CG coordinates in the CGMD model compared with the Full MD model. (a) The radius distribution function $g(r)$ of the CGCOM model. (b) The distribution of the pairwise distance $g'(r)$  between two CG coordinates within the same molecular of the CG3 model. (c)-(j) The distribution of distance $g'(r)$   between two CG coordinate (1-2,1-3,1-4,2-3,2-4,3-4, respectively) within the same molecular of CG4 model.}
    \label{fig:app_rho}
\end{figure}

\begin{figure}
    \centering
    \includegraphics[width=0.3\textwidth]{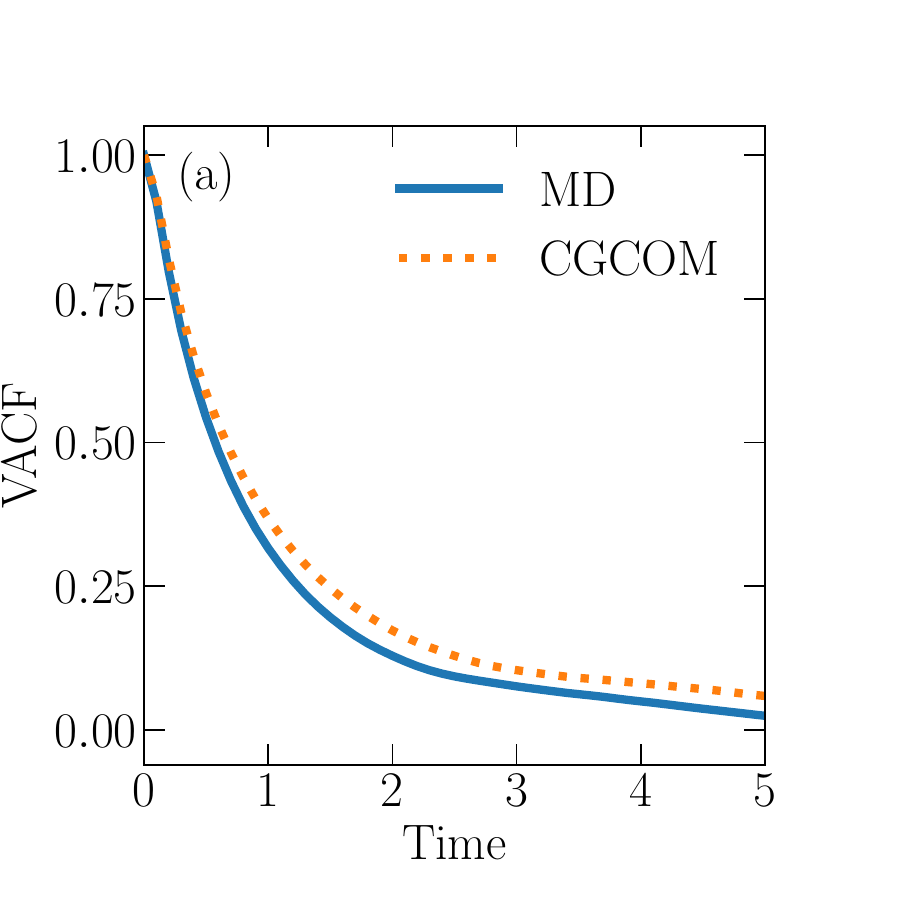}
    \includegraphics[width=0.3\textwidth]{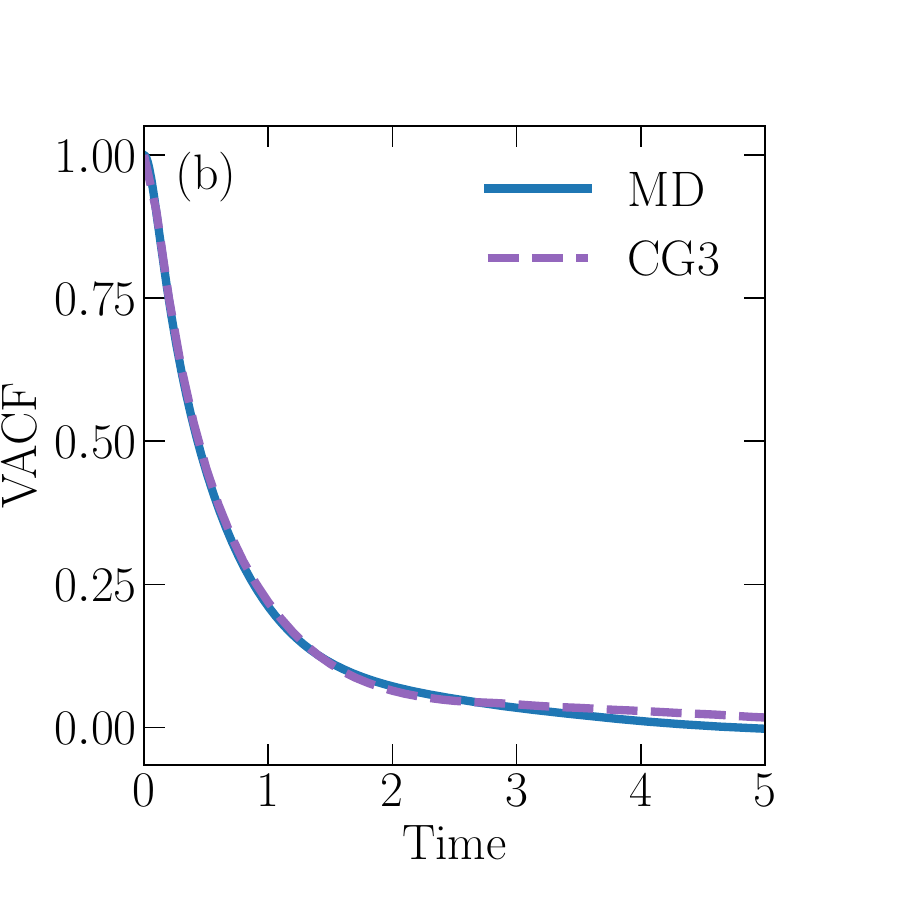}
    \includegraphics[width=0.3\textwidth]{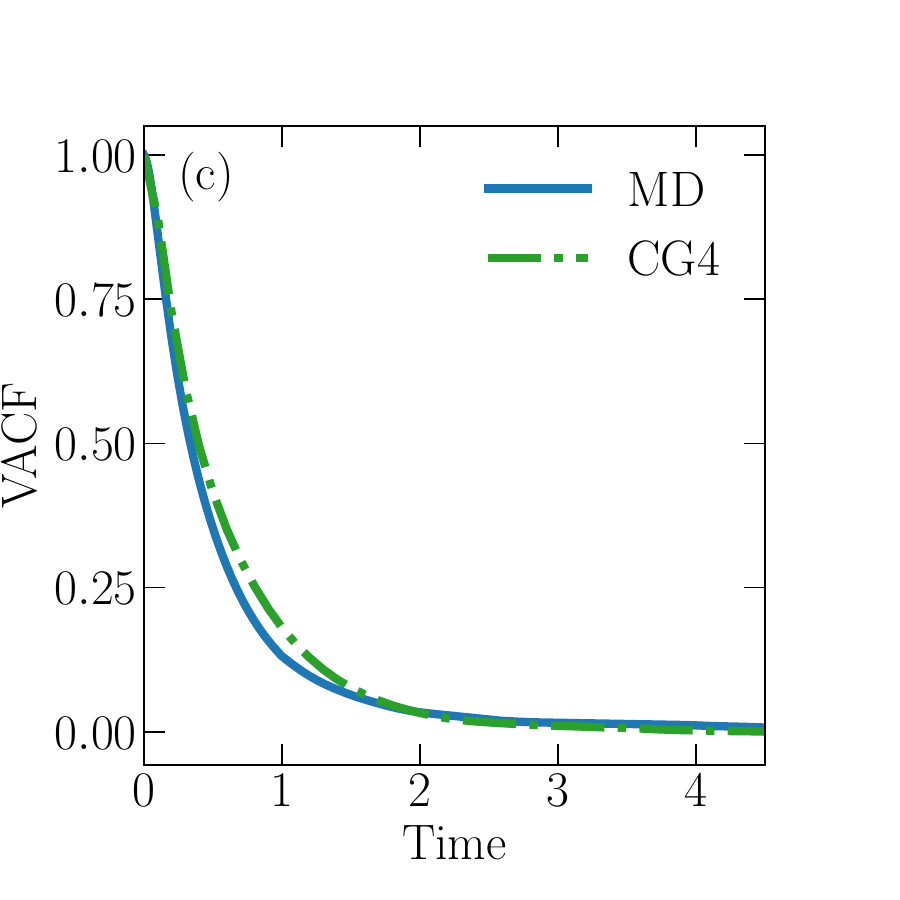}
    \includegraphics[width=0.3\textwidth]{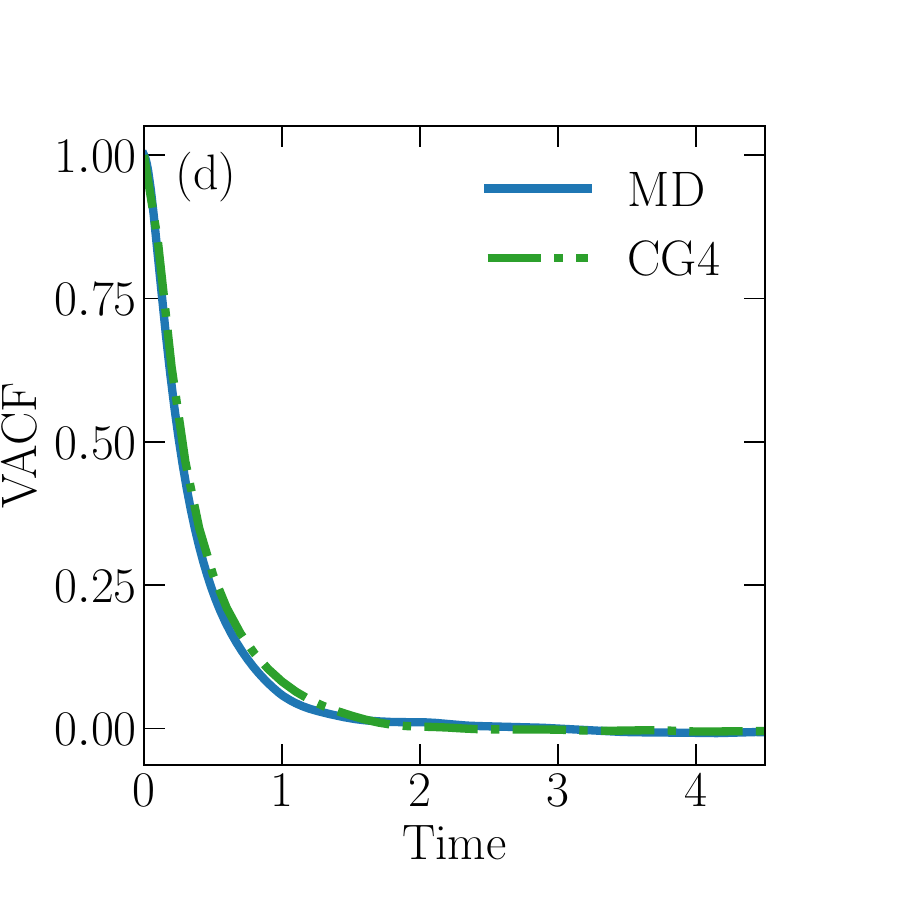}
    \includegraphics[width=0.3\textwidth]{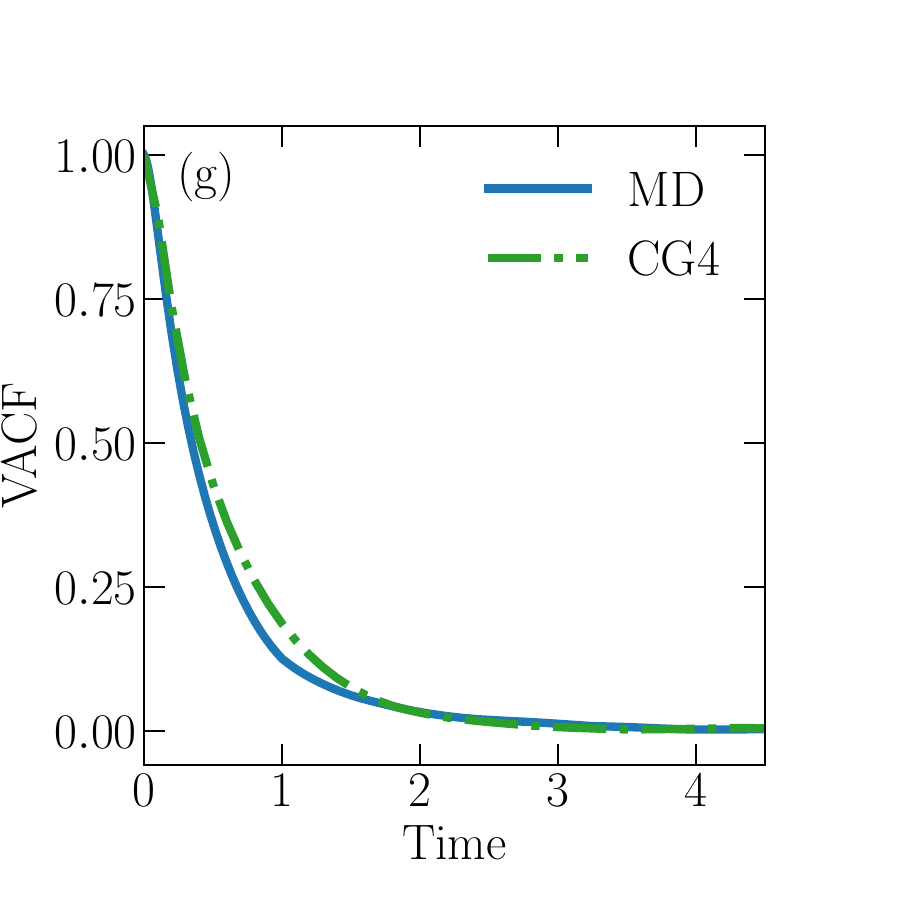}
    \includegraphics[width=0.3\textwidth]{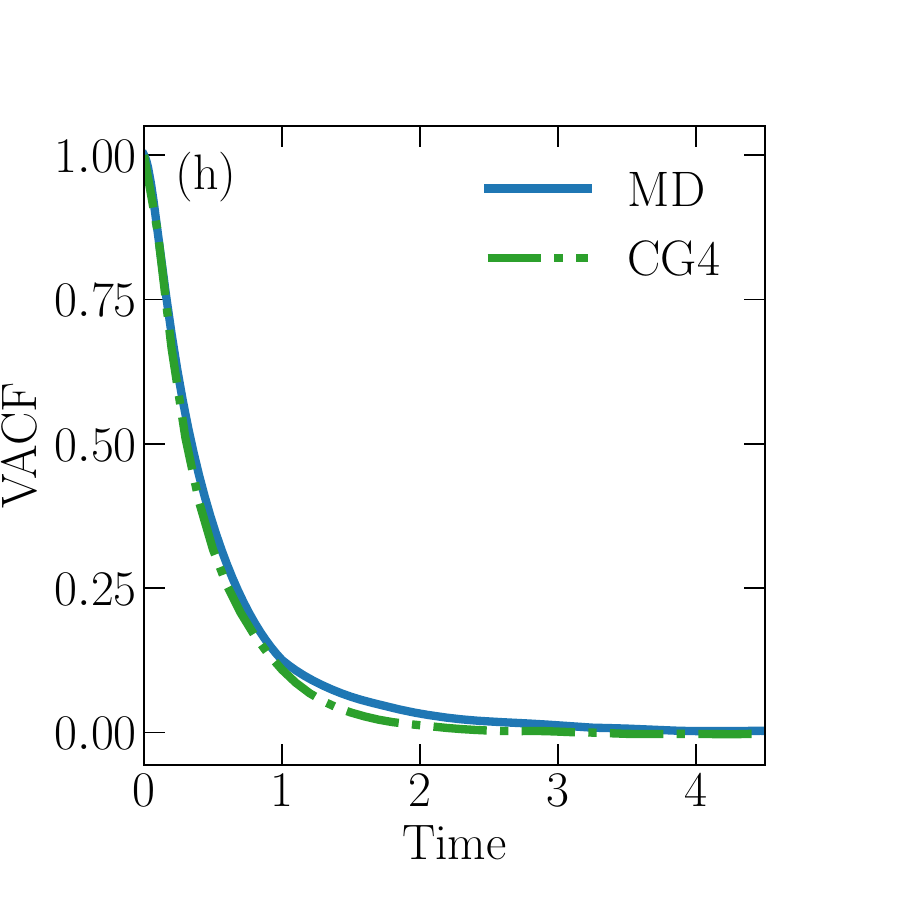}
    \caption{The VACF of CVs in different CG models and their comparison with Full MD. (a) VACF of CVs in CG3 model and the comparison with Full MD. (b)-(g) VACF of CVs (1,2,3,4, respectively) in CG4 model and the comparison with Full MD.}
    \label{fig:app_vacf}
\end{figure}

To establish a fair comparison of the various CG models, we examine the static and dynamic properties that have been widely used as benchmark problems for model validation. Fig. \ref{fig:app_rho} shows the radial distribution function between the CG coordinates. For all three CG models, their predictions agree well with the full MD results, which verifies the accuracy of the CG free energy function $U(\mb Q)$.  Fig. \ref{fig:app_vacf} shows the normalized velocity auto-correlation function (VACF) of the CG variables. Similar to the static properties, the prediction of the CG models agrees well with the full MD results, which verifies the accuracy of the many-body memory term $\mb K(\mb Q, t)$.

Despite of the good agreement, we emphasize that these properties are essentially defined under the marginal density near equilibrium. The applicability to non-equilibrium processes is generally unwarranted, and further relies on the consistency in the conditional probability density function associated with the CG projection operator (i.e., Eq. \eqref{eq:conditional_prob_metric}) as discussed in Sec. \ref{sec:method}.

\section{CG variable training}
{
In this section, we show that Eq. \ref{eq:W_constraint} does not impose further constraints on constructing the CG variables. We examine the equivalence of two CG mapping. For instance, when the centers of mass (COMs) are utilized as the only CG variables, the information they encompass is the same as the COMs plus a constant or the COMs multiplied by a constant.
\begin{definition}
For two CG map $\phi$ and $\check{\phi}$, if there exist a map $\mathbb T$, such that $\phi(\mathbf q, \mathbf p) = \mathbb T (\check{\phi}(\mathbf q, \mathbf p))$, we say $\phi \leq \check{\phi}$. Also, if $\phi \leq \check{\phi}$ and $\check{\phi}\leq \phi$, we say that the two maps are equivalent.
\end{definition}
In other words, if $\phi \leq \check{\phi}$, any properties that can be measured from the CG coordinate $\phi$, denote as $\mathcal{F}(\phi)$ , can also be meansured from $\check{\phi}$ as  $\mathcal{F}(\mathbb T (\check{\phi}))$. If $\phi \leq \check{\phi}$ and a CG model $\frac{\intd}{\intd t}{\check\phi}  = \check{\mathcal{L}} \check{\phi}$ is constructed using CG coordinate $\check{\phi}$, the CG model for $\phi$ can be constructed by the chain rule $\frac{\intd }{\intd t }\phi =  \frac{\delta T}{\delta \check{\phi}} \check{\mathcal{L}} \check{\phi}$ if $\check{\mathcal{L}}$ is deterministic or It\^o's lemma if $\check{\mathcal{L}}$ includes stochastic term.

\begin{proposition}
For any linear CG map $\phi$ defined by a matrix $\mb W\in \mathbb R^{N_m\times (m-1) }$, there exists a map $\check\phi$, defined by a matrix $\check{\mb W} \in \mathbb R^{N_m\times m }$, subject to the following constraints
\begin{equation}
0 < \check{w}_{ni} < 1, \quad  \sum_{i=1}^m \check{w}_{ni} = 1.
\label{eq:app_W_constraint}
\end{equation}

\begin{proof}
    For any $\mb W$ with entry $w_{ij}$, $\check{\mb W}$ can be constructed as 
    \begin{equation}
        \check w_{ni} = \left\{ 
        \begin{aligned}
      &\frac{w^1_{ni}-  \tilde{w}^1}{\bar{w}}, & 0< i < m,\\
    &1-\sum_{i=1}^{m-1} \check  w _{ni}, & i = m, 
        \end{aligned}
        \right.
    \end{equation}
where $\tilde{w} = \min_{n,i} w_{ni}$ and $\bar{w} = \max_i\sum_{i}^{m-1} (w_{ni} -\tilde{w}) $. It's obvious that $\check{w}_{ni} \in [0,1]$ for $0<i<m$. Since 
$$
\sum_{i=1}^{m-1} \check{w}_{ni} =  \frac{\sum_{i=1}^{m-1} (w_{ni}-  \tilde{w})}{\bar{w}} \in [0,1],
$$ 
we have $\check{w}_{nm} \in [0,1]$. Also, $\sum_i \check{w}_{ni}=1$ for every $n$, hence $\check{W}$ satisfies the constraints. Since
\begin{equation}
    \sum_i^m (\sum_n^{N_m} \check{w}_{ni}) \check{\mathbf Q}_i^I = \sum_i^m  \sum_n^{N_m} \check{w}_{ni}\mathbf q^I_n = \sum_n^{N_m} \mathbf q^I_n,
\end{equation}
we have 
\begin{equation}
   \begin{aligned}
\mathbf  Q^I_{i} &= \frac{\sum_n^{N_m} w_{ni}\mathbf q_{I,n} }  {\sum_n^{N_m}w_{ni}} \\
& =  \frac{\sum_n^{N_m} (\bar{w} \check{w}_{ni}+\tilde{w})\mathbf q_{I,n} }  {\sum_n^{N_m}(\bar{w}  \check{w}_{ni}+\tilde{w})}\\
& = \frac{\bar{w}( \sum_n^{N_m} \check{w} _{ni}) \check{\mathbf Q}^I_{i}+ \tilde{w}\sum_j^m\left(\sum_n^{N_m} \check{w}_{nj} \right) \check{\mathbf Q}^I_{j} }{\sum_n^{N_m}(\bar{w}  \check{w}_{ni}+\tilde{w})}.
\end{aligned} 
\end{equation}
Therefore,  $\mathbf Q^I$ is a linear transformation of $\check{\mathbf Q}^{I}$ for all $I=1,\cdots,M$. Similarly, $\mathbf P^I$ can be shown as a linear transformation of $\check{\mathbf P}^I$.
\end{proof}
\end{proposition}
}


\clearpage

\end{document}